\documentclass{aa}  
\usepackage{dcolumn}

\usepackage{graphicx}
\usepackage{txfonts}
\newcommand{\Ha}{H$\alpha$}
\newcommand{\Hb}{H$\beta$}

\newcommand{\Lya}{Ly$\alpha$}
\newcommand{\sVl}[3]{#1\,{\sc #2}]\,$\lambda{#3}$}
\newcommand{\Nl}[3]{#1\,{\sc #2}\,$\lambda{#3}$}
\makeatletter
 \def\hlinewd#1{%
   \noalign{\ifnum0=`}\fi\hrule \@height #1 \futurelet
    \reserved@a\@xhline}
\makeatother
\newcommand{\htopline}{\hlinewd{.8pt}}
\newcommand{\hmidline}{\hlinewd{.2pt}}
\newcommand{\hbotline}{\htopline}
\newcommand{\mcc}[1]{\multicolumn{1}{c}{#1}}

\newcommand{\mcr}[1]{\multicolumn{1}{r}{#1}}
\newcommand{\kms}{km\,s$^{-1}$}
\DeclareMathOperator{\log10}{log10}

\begin{document} 

    \title{Vertical broad-line region structure in nearby active galactic
          nuclei} 

   \author{W. Kollatschny 
          ,  M. Zetzl 
          }

   \institute{Institut f\"ur Astrophysik, Universit\"at G\"ottingen,
              Friedrich-Hund Platz 1, D-37077 G\"ottingen, Germany\\
              \email{wkollat@astro.physik.uni-goettingen.de}
}

   \date{Received April 12, 2013; accepted July 18, 2013}
   \authorrunning{Kollatschny, Zetzl}
   \titlerunning{Vertical broad-line region structure}

 
  \abstract
   {Broad emission lines are emitted
in the surroundings of supermassive black holes in the centers
 of active galactic nuclei (AGN).
   This region is spatially not resolved.}
   {We intend to get information on the structure and geometry
  of this broad emitting line region (BLR) based
    on line profile observations.}
   {We model the rotational and turbulent velocities in the
   line-emitting regions based on observed full-width at half maximum
   line values
   (FWHM) and $\sigma_{\mathrm{line}}$ of the variable
    broad emission lines in  four nearby AGN: NGC\,3783, NGC\,7469, NGC\,5548,
    and 3C~390.3.
    On the basis of these velocities, we estimate the height
    of the line-emitting regions above the midplane
    in context with their distances
    from the center.}
   {The H$\beta$ lines are emitted in a more flattened configuration
above the midplane in comparison to the highly ionized lines. 
The H$\beta$ lines originate at heights
of 0.7 to 1.6 light-days and at distances of 1.4 to 24 light-days
with height/distance ($H/R$) ratios of only 0.07 to 0.5.
The highly ionized lines originate at smaller radii than the H$\beta$ lines
and/or at greater distances above the midplane with $H/R$ values of 0.2 to
1.7. In total, the emission lines do not originate in a thin
atmosphere of an accretion disk but rather at very extended regions
above an accretion disk.
The observed geometries of the line-emitting regions resemble
the geometries of accretion disk wind models.
Furthermore, the angle of the central opening cone (generated by the
emitting regions of the highly ionized lines)
is small for those galaxies with slow rotational velocities and increases
with the rotation velocity of the central region.} 
   {The derived geometries
   of the line-emitting regions of all four AGN
   are consistent with the geometries that are predicted in outflowing
   disk wind models.}

   \maketitle
%

\section{Introduction}
It is now generally accepted that the emitted continuum flux
in active galactic nuclei (AGN)
is generated by accretion of matter onto a supermassive black hole.
This central black hole is surrounded by an accretion
disk (e.g., Lynden-Bell\citealt{lynden69}).
The broad optical/UV emission lines we see in the spectra of
Seyfert~1 galaxies and/or quasars
are caused by photoionization of the central ionizing source
at distances of about 1 to 100 light-days from the center.
These broad emitting line regions are spatially unresolved,
even for the nearest AGN.

First pieces of information about the distances
 of the broad-line emitting
regions from the central ionizing region
have been obtained 
from the delayed variability of the integrated emission line intensities
 with respect
to the ionizing continuum: the so-called reverberation mapping method
(e.g., Cherepashchuk \& Lytyi\citealt{cherep73},
 Clavel et al.\citealt{clavel91}).
Furthermore, the broad-line region is stratified. The highly
ionized lines originate closer to the central ionizing source than the
lower ionized lines (e.g., Gaskell \& Sparke\citealt{gaskell86},
Krolik et al.\citealt{krolik91},
Korista et al.\citealt{korista95}, Peterson \& Wandel \citealt{peterson99},
 Kollatschny\citealt{kollatschny01}).

 Individual delays of emission
line segments (velocity delay maps) could be verified in a few cases.
It has been shown that the broad-line region (BLR)
 is gravitationally bound by means of
the  velocity-resolved reverberation mapping method
(e.g., Gaskell\citealt{gaskell88}, Koratkar \&
 Gaskell\citealt{koratkar89},\citealt{koratkar91a},\citealt{koratkar91b})
 with a slight inflow
(see Gaskell \& Goosmann\citealt{gaskell13}).
Other studies found
signatures of accretion disk
winds (Kollatschny\citealt{kollatschny03}, Bentz et al.\citealt{bentz10},
Wang et al.\citealt{wang11}).

For a general
review of our present knowledge of the structure of the BLR see
Gaskell\cite{gaskell09}.
There exist various models regarding the geometry and structure of both
 accretion disks in AGN and accretion disk winds
(e.g., Osterbrock\citealt{osterbrock78}; Blandford\citealt{blandford82};
Collin-Souffrin et al.\citealt{collin88};
 Emmering et al.\citealt{emmering92};  
 K\"onigl \& Kartje\citealt{koenigl94}; DeKool \& Begelman\citealt{dekool95};
 Murray \& Chiang\citealt{murray97},
\citealt{murray98}; Bottorff et al.\citealt{bottorff97}; 
Blandford \& Begelman\citealt{blandford99};
Elvis\citealt{elvis00};
Proga, Stone \& Kallman\citealt{proga00}; Proga \& Kallman\citealt{proga04};
 Kollatschny\citealt{kollatschny03},\citealt{kollatschny13a};
Everett\citealt{everett05};
Ho\citealt{ho08}; Gaskell\citealt{gaskell10};
 Goad et al.\citealt{goad12}; Flohic et al.\citealt{flohic12}
and references therein).
More direct evidence for a flattened BLR has come from the following three
lines of evidence:
(1) Wills \& Browne\cite{wills86} showed that the FWHM of
H$\beta$ is strongly correlated
with the orientation estimated from radio observations.
(2) The reverberation mapping transfer functions of the low-ionization lines
(Krolik et al.\citealt{krolik91}; Horne et al,\citealt{horne91}) showed that
 there was
little or no material on the line of sight and were consistent with a flattened
disk.
(3) Energy-budget requirements plus the absence of BLR absorption require the
BLR to be flattened (Gaskell et al.\citealt{gaskell07};
Gaskell\citealt{gaskell09}).

The origin of the accretion disk winds is explained by radiation-driven
winds or magnetocentrifugal winds.
However, many details of the vertical broad-line region structure are still
unknown.

The profiles of the broad emission lines in AGN can be parameterized by the
ratio of
their full-width at half maximum (FWHM) to their line dispersion
$\sigma_{\mathrm{line}}$. We demonstrated 
in two recent papers (Kollatschny \& Zetzl\citealt{kollatschny11},
hereafter called Paper I, and Kollatschny \& Zetzl\citealt{kollatschny13a},
hereafter called Paper II)
that there exist general trends between 
the line width (FWHM) and
 the line width ratio 
FWHM/$\sigma_{\mathrm{line}}$ in these broad emission lines.
Different emission lines exhibit different systematics in the 
FWHM/$\sigma_{\mathrm{line}}$-vs-FWHM diagram. 
 The line width FWHM reflects the
rotational motion of the broad-line gas in combination with associated
turbulent motions. These turbulent velocities in the line-emitting regions are
different for different emission lines.
The rotational and turbulent velocities
give us information on the accretion disk
height with respect to the accretion disk radius of the line-emitting regions.
We know the absolute numbers of the line-emitting radii from reverberation
mapping. Therefore, one can get information on the absolute heights of
 the line-emitting regions
above the accretion disks.
In a third paper
(Kollatschny \& Zetzl\citealt{kollatschny13b},
hereafter called Paper III), we presented first results with regard to
 the broad-line region
 geometry of NGC~5548 based on their line profiles.

Here we present results on the broad-line region
geometries of three new galaxies:
NGC~3783, NGC~7469, and 3C~390.3.
We know the profile parameters of at least four different emission lines
in these galaxies, and we know in addition the distances of their line-emitting
regions from the center based on former variability campaigns. 
On the basis of that information
we are able to make statements about the
geometry and structure of broad emission line regions
in these AGN, which are emitting spectra with
diverse line widths.


\section{The data sample}


The current investigation is based on the same sample of
variable AGN spectra
of Peterson et al.\cite{peterson04} as in Papers I to III.
The original AGN sample consisted of 37 galaxies. 
Variable broad emission lines of H$\alpha$, H$\beta$,
 Ly$\alpha$, \ion{He}{ii}\,$\lambda 4686$,  \ion{He}{ii}\,$\lambda 1640$,
\ion{C}{iii}]\,$\lambda 1909$, \ion{C}{iv}\,$\lambda 1550$, and 
\ion{Si}{iv}+\ion{O}{iv]}\,$\lambda 1400$ lines have been
obtained with different ground-based telescopes, while the
UV spectra were taken with the International Ultraviolet Explorer (IUE) and/or
the Hubble Space Telescope during dedicated variability campaigns.
This sample has the advantage that all the spectra were reduced
in exactly the same way and that the spectra of each
galaxy were intercalibrated with respect to each other.

For our analysis we use the mean as well as the
root-mean-square (rms) profiles of the variable broad emission lines.
Usually the observed broad-line profiles in AGN are more or less
 contaminated by
additional narrow emission line components from the narrow line region.
To avoid any major ambiguity we inspected only the
rms line profiles out of the sample of 
Peterson\cite{peterson04} (see Papers I to III).
 The rms profiles display the clean profiles of
 the variable broad emission lines.
 The narrow line components disappear in these
 spectra as they are constant over timescales of years. The
 emission line profiles in our AGN sample can be
 parameterized by their line widths FWHM and $\sigma_{\mathrm{line}}$.
The relationship between
FWHM and $\sigma_{\mathrm{line}}$ contains 
information on the shape of the profile.

In Paper III we presented results on the broad-line region structure in
NGC~5548 based on their optical/UV broad-line profiles. The optical/UV
rms emission line profiles were based on two separated variability campaigns,
while the  H$\beta$  line has been monitored over a period of 14 years.
We use these data as comparison for the present investigation of the
broad-line region structure of three further AGN.   
For the current investigation we selected all those AGN
in the study of Peterson et al.\cite{peterson04}
 where 
line profile data of at least four emission lines have
been published: NGC~3783, NGC~7469, 3C~390.3.

The galaxy NGC~7469 has been the target of a large
 international variability campaign
with ground-based telescopes in the optical and with the
IUE satellite for the UV spectral
 lines.
In 1996 from June 2 to July 30, 
 optical spectrophotometric observations
of NGC 7469 were completed (Collier et al.\citealt{collier98}) on 54 nights.
From 1996 June 10 to July 29, the IUE satellite monitored
the Seyfert 1 galaxy NGC 7469 continuously in an attempt to measure time delays
between the UV continuum and the broad UV emission-line fluxes 
(Wanders et al.\citealt{wanders97}). 

The galaxy NGC~3783 has been the target of a large
 international variability campaign
with ground-based telescopes in the optical and with IUE for the UV spectral
 lines.
The Seyfert 1 galaxy NGC~3783 was monitored in the optical
from 1991 December through 1992 August (Stirpe et al.\citealt{stirpe94}).
It was observed with IUE at 69 epochs between 1991 December 21 and
 1992 July 29 (Reichert et al.\citealt{reichert94}).

The galaxy 3C~390.3 was monitored in a  ground-based optical monitoring campaign
from 1994 October through 1995 October (Dietrich et al.\citealt{dietrich98}).
In parallel an extensive monitoring campaign was carried out
with the IUE
satellite from 1994 December 31 to 1996 March 5 
 (O'Brien et al.\citealt{obrien98}).
For the values of the line widths in 3C~390.3 (Table 1, Fig.~3) we used the
unblended halfs of the emission lines (Peterson et al.\citealt{peterson04}).


\section{Results}


We parameterize the observed rms line profiles in our galaxies
 by their line width FWHM as well as by the ratio of
their FWHM to their line dispersion
$\sigma_{\mathrm{line}}$.   
 This gives us information on the ratio of the accretion disk
height with respect to the accretion disk radius of the line-emitting region.
As we know the absolute numbers of the line-emitting radii from reverberation
mapping studies, we are additionally 
able to make statements about the
geometry and height of the broad emitting line regions above the midplane
in different AGN.
Here we present results for the three galaxies
NGC~3783, NGC~7469, and 3C~390.3 and compare these findings with those of
NGC~5548.

All spectral data we use in the current investigation are listed
in Table 1.  The H$\beta$ line widths FWHM in our galaxies range from 2000 to
10\,000  km\,s$^{-1}$, the widths of  \ion{C}{iv}\,$\lambda 1549$ range from
4300 to 9000  km\,s$^{-1}$.

\begin{table*}[htbp]
    \centering
       \leavevmode
       \tabcolsep0.3mm 
        \newcolumntype{d}{D{.}{.}{-2}} 
        \newcolumntype{p}{D{+}{\,\pm\,}{-1}}
        \newcolumntype{K}{D{,}{}{-2}}
\caption{Line profile parameters  as well as
radii and heights above the midplane of the line-emitting regions
of individual emission lines.
}
\begin{tabular}{lKKKKKKdKd}
 \htopline
\hspace{3mm} Line &\mcr{FWHM }& \mcr{FWHM/$\sigma$} &\mcc{$v_{\text{turb}}$}&\mcc{$v_{\text{rot}}$} &\mcc{Radius}
                  & \mcc{Height} & \mcr{$H/R$} & \mcc{Height$_{\text{corr}}$} & \mcc{$H_{\text{corr}}/R$}\\
\hspace{3mm}      &\mcr{[\kms{}]}&&\mcc{[\kms{}]}&\mcc{[\kms{}]} &\mcc{[ld]}
                  & \mcc{[ld]} & \mcr{} & \mcc{[ld]} & \mcr{}\\
 \hmidline
\multicolumn{3}{l}{\hspace{-3mm}NGC 7469}\\
 \hmidline
\hspace{3mm}\Nl{He}{ii}{1640}                              &10725,\pm{1697}&2.88,\pm{0.46~}&1063,^{+1147}_{-1024} &6281,^{+949}_{-1080~~}&0.6,^{+0.3}_{-0.4}&0.1,\pm{0.1}&0.17&0.2,\pm{0.2}&0.33\\           
\hspace{3mm}\ion{Si}{iv}+\ion{O}{iv]}\,{$\lambda$1400}     &6033,\pm{1112}&1.73,\pm{0.34}&2745,^{+1267}_{-1070} &3185,^{+941}_{-1524}&1.7,^{+0.3}_{-0.3}&1.5,\pm{1.0}&0.88&1.1,\pm{0.9}&0.65\\               
\hspace{3mm}\Nl{C}{iv}{1549}                               &4305,\pm{422}&1.64,\pm{0.18}&1462,^{+297}_{-285} &2406,^{+314}_{-354}&2.5,^{+0.3}_{-0.2}&1.5,\pm{0.4}&0.60&3.0,\pm{0.7}&1.20\\                   
\hspace{3mm}\Hb                                            &2169,\pm{459}&1.49,\pm{0.38}&447,^{+220}_{-198} &1259,^{+280}_{-325}&4.5,^{+0.7}_{-0.8}&1.6,\pm{0.9}&0.36&1.4,\pm{0.9}&0.31\\                    
\hspace{3mm}\Ha                                            &1615,\pm{119}&1.39,\pm{0.13}&292,^{+52}_{-49} &941,^{+73}_{-77}&4.7,^{+1.6}_{-1.3}&1.5,\pm{0.6}&0.32&3.5,\pm{1.3}&0.74\\                     
 \hmidline
\multicolumn{3}{l}{\hspace{-3mm}NGC 3783}\\
\hmidline
\hspace{3mm}\Nl{He}{ii}{1640}                              &8008,\pm{1268}&2.07,\pm{0.34}&2859,^{+1262}_{-1094} &4445,^{+955}_{-1303}&1.4,^{+0.8}_{-0.5}&0.9,\pm{0.7}&0.64&0.7,\pm{0.6}&0.50\\     
\hspace{3mm}\ion{Si}{iv}+\ion{O}{iv]}\,{$\lambda$1400}     &6343,\pm{2021}&1.82,\pm{0.59}&2599,^{+1241}_{-1226} &3435,^{+1450}_{-2284}&2.0,^{+0.9}_{-1.1}&1.5,\pm{1.5}&0.75&1.2,\pm{1.3}&0.60\\    
\hspace{3mm}\Nl{C}{iv}{1549}                               &3691,\pm{475}&1.25,\pm{0.17}&2203,^{+448}_{-419} &1746,^{+487}_{-686}&3.8,^{+1.0}_{-0.9}&4.8,\pm{2.5}&1.26&6.3,\pm{3.1}&1.66\\         
\hspace{3mm}\Hb                                            &3093,\pm{529}&1.76,\pm{0.33}&591,^{+216}_{-210} &1800,^{+321}_{-354}&10.2,^{+3.3}_{-2.3}&3.3,\pm{1.8}&0.32&2.3,\pm{1.5}&0.23\\         
\hmidline
\multicolumn{3}{l}{\hspace{-3mm}3C 390.3}\\
\hmidline
\hspace{3mm}\Hb                                            &9958,\pm{1046}&3.21,\pm{0.35}&323,^{+559}_{-10322} &5793,^{+562}_{-583}&23.6,^{+6.2}_{-6.7}&1.3,\pm{42.1}&0.06&1.6,\pm{42.1}&0.07\\              
\hspace{3mm}\Nl{He}{ii}{4686}                              &8488,\pm{1842}&2.57,\pm{0.59}&1277,^{+1334}_{-1158} &4964,^{+1083}_{-1334}&27.2,^{+31.2}_{-24.8}&7.0,\pm{11.0}&0.26&4.9,\pm{9.3}&0.18\\        
\hspace{3mm}\Nl{C}{iv}{1549}                               &8989,\pm{2987}&2.04,\pm{0.68}&4103,^{+5896}_{-2630} &4742,^{+2274}_{-14741}&35.7,^{+11.4}_{-14.6}&30.9,\pm{106.5}&0.87&21.8,\pm{81.6}&0.61\\   
\hspace{3mm}\Lya                                           &8732,\pm{985}&2.21,\pm{0.27}&2696,^{+1288}_{-1063} &4938,^{+735}_{-1007}&58.6,^{+27.7}_{-27.2}&32.0,\pm{22.5}&0.55&45.1,\pm{27.8}&0.77\\       
\hbotline
\end{tabular}
\end{table*}

\subsection{Observed and modeled emission
line width ratios}

We present in Figs.\,1 to 3 the observed line widths and line width ratios
of the emission lines in NGC~7469, NGC~3783, 3C~390.3 as well as the
related turbulent $v_{\mathrm{turb}}$ and 
rotational velocities $v_{\mathrm{rot}}$ of the line-emitting regions
(see Papers I to III). It has been demonstrated by Goad et al.(\citealt{goad12})
(see Paper II) that turbulent motions in the accretion disk produce
Lorentzian profiles.

We have shown in Papers I and II that
the ratio of the turbulent velocity $v_{\mathrm{turb}}$
in the line-emitting region
 with respect to the rotational velocity $v_{\mathrm{rot}}$ 
gives us information on the ratio of the accretion disk
height $H$ with respect to the accretion disk radius $R$
 of the line-emitting regions.
By knowing the distances $R$
of the line-emitting regions based on earlier reverberation
mapping studies we are able to estimate the heights $H$ of the line-emitting 
regions. We give in Table 1 the heights of
the line-emitting regions in units of light-days as well as the
ratio $H/R$ for the individual emission lines in our galaxies.
%
%
\begin{figure}
\includegraphics[width=6.2cm,angle=270]{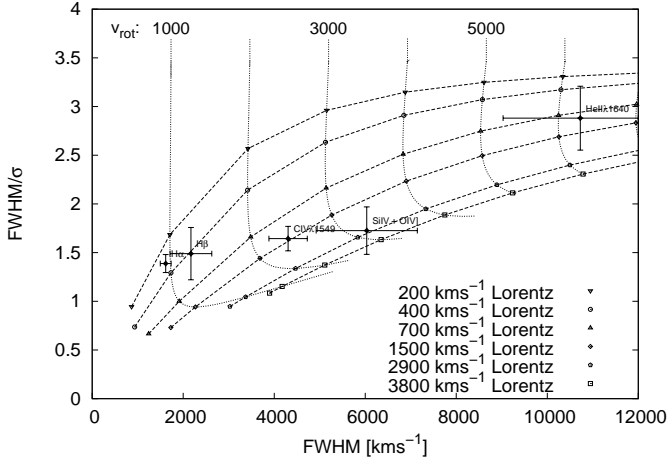} 
       \vspace*{5mm} 
  \caption{NGC7469:
Observed and modeled line width ratios
 FWHM/ $\sigma_{\mathrm{line}}$ versus line width FWHM. The dashed curves
 represent the corresponding theoretical
line width ratios based on rotational line broadened
 Lorentzian profiles (FWHM = 200 to
3800\,$km\,s^{-1}$). The rotation velocities go from 1000 to 6000\,$km\,s^{-1}$
(curved dotted lines from left to right).}
   \label{ngc7469.ps}
\end{figure}
%
%
%
\begin{figure}
\includegraphics[width=6.2cm,angle=270]{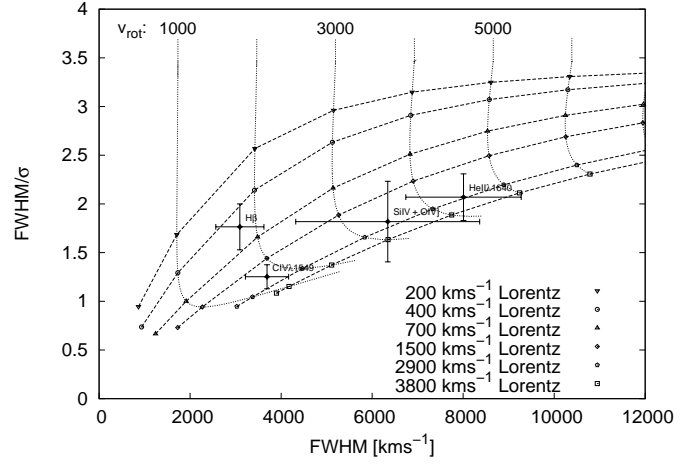} 
       \vspace*{5mm} 
  \caption{NGC3783: Observed and modeled
 line width ratios
 FWHM/ $\sigma_{\mathrm{line}}$ versus line width FWHM.}
   \label{ngc3783.ps}
\end{figure}
%
%
%
%
\begin{figure}
\includegraphics[width=6.2cm,angle=270]{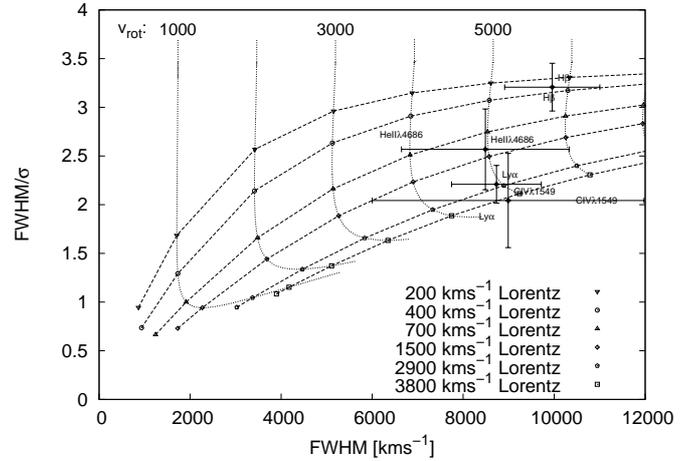} 
       \vspace*{5mm} 
  \caption{3C~390.3: Observed and modeled line width ratios
 FWHM/ $\sigma_{\mathrm{line}}$ versus line width FWHM.}
   \label{3c390.ps}
\end{figure}
%
%


We demonstrated in Papers I and II that dedicated turbulent velocities 
belong to individual emission lines:
We showed that there exist general trends between 
the line width FWHM and the line width ratio 
FWHM/$\sigma_{line}$ of the individual broad emission lines.
Different emission lines exhibit different systematics in this diagram.
In a second step we modeled the observed trends. We could show that the 
shapes of the line profiles are caused by two dominant parameters:
rotation and turbulent velocities. These turbulent velocities are different
for the different emission lines.
Additional small deviations from these general trends
can be explained
by additional inflow/outflow motions  causing line asymmetries in 
individual galaxies, orientation effects of the line-emitting regions,
optical thickness effects, etc.

\begin{table*}[htbp]
    \centering
       \leavevmode
       \tabcolsep4mm 
        \newcolumntype{d}{D{.}{.}{-2}} 
        \newcolumntype{p}{D{+}{\,\pm\,}{-1}}
        \newcolumntype{K}{D{,}{}{-2}}
\caption{Radii and heights of the line-emitting regions
in units of the Schwarzschild radii
of their central black holes for the galaxies NGC~7469, NGC~3783, and 3C390.3.
These data are also given for the two variability campaigns of NGC~5548
during 1988/89 and 1993, as well as for the H$\beta$ monitoring campaign
of NGC~5548 over an additional period of 11 years (see Paper III).
 Furthermore, the optical
continuum luminosities at 5100 \AA{}  are given for the individual
campaigns.
}
\begin{tabular}{lKKdKd|c}
 \htopline
\hspace{3mm} Line &\mcc{Radius} & \mcc{Height} & \mcr{$H/R$} & \mcc{Height$_{\text{corr}}$} & \mcc{$H_{\text{corr}}/R$}&$\log10(\lambda{}L_{\lambda})$\\
\hspace{3mm}      &\mcc{[R$_{\mathrm{S}}$]}   & \mcc{[R$_{\mathrm{S}}$]} & \mcr{} & \mcc{[R$_{\mathrm{S}}$]} & \mcr{}&\mcr{[erg\,s$^{-1}$]}\\
 \hmidline
\multicolumn{2}{l}{\hspace{-3mm}NGC 7469} &\multicolumn{4}{l}{ ($M_{\mathrm{BH}} = 12.2\times10^{6}\, \mathrm{M}_{\odot}$, $R_{\mathrm{S}} = 1.39\times10^{-3}\,\mathrm{ld} = 3.60\times10^{12}\,\mathrm{cm}$)}\\
 \hmidline
\hspace{3mm}\Nl{He}{ii}{1640}                           &431,^{+215}_{-287}&71,\pm{71}&0.17&143,\pm{143}&0.33&43.56$\,\pm{\,0.10}$\\         
\hspace{3mm}\ion{Si}{iv}+\ion{O}{iv]}\,{$\lambda$1400}  &1221,^{+215}_{-215}&1078,\pm{718}&0.88&790,\pm{646}&0.65&$\cdot{}$\\  
\hspace{3mm}\Nl{C}{iv}{1549}                            &1796,^{+215}_{-143}&1078,\pm{287}&0.60&2156,\pm{503}&1.20&$\cdot{}$\\ 
\hspace{3mm}\Hb                                         &3234,^{+503}_{-575}&1150,\pm{646}&0.36&1006,\pm{646}&0.31&$\cdot{}$\\ 
\hspace{3mm}\Ha                                         &3378,^{+1150}_{-934}&1078,\pm{431}&0.32&2515,\pm{934}&0.74&$\cdot{}$\\
 \hmidline
\multicolumn{2}{l}{\hspace{-3mm}NGC 3783} &\multicolumn{4}{l}{($M_{\mathrm{BH}} = 29.8\times10^{6}\,  \mathrm{M}_{\odot}$, $R_{\mathrm{S}} = 3.40\times10^{-3}\,\mathrm{ld} =  8.80\times10^{12}\,\mathrm{cm} $)}\\
\hmidline
\hspace{3mm}\Nl{He}{ii}{1640}                           &411,^{+235}_{-147}&264,\pm{205}&0.64&205,\pm{176}&0.50&42.55$\,\pm{\,0.18}$\\    
\hspace{3mm}\ion{Si}{iv}+\ion{O}{iv]}\,{$\lambda$1400}  &588,^{+264}_{-323}&441,\pm{441}&0.75&353,\pm{382}&0.60&$\cdot{}$\\    
\hspace{3mm}\Nl{C}{iv}{1549}                            &1118,^{+294}_{-264}&1412,\pm{735}&1.26&1853,\pm{912}&1.66&$\cdot{}$\\ 
\hspace{3mm}\Hb                                         &3001,^{+971}_{-676}&971,\pm{529}&0.32&676,\pm{441}&0.23&$\cdot{}$\\   
\hmidline
\multicolumn{2}{l}{\hspace{-3mm}3C 390.3}  &\multicolumn{4}{l}{($M_{\mathrm{BH}} = 287\times10^{6}\,  \mathrm{M}_{\odot}$, $R_{\mathrm{S}} = 32.7\times10^{-3}\,\mathrm{ld} = 84.77\times10^{12}\,\mathrm{cm}$)}\\
\hmidline
\hspace{3mm}\Hb                                         &721,^{+189}_{-204}&39,\pm{1286}&0.06&48,\pm{1286}&0.07&43.62$\,\pm{\,0.07}$\\     
\hspace{3mm}\Nl{He}{ii}{4686}                           &831,^{+953}_{-757}&213,\pm{336}&0.26&149,\pm{284}&0.18&$\cdot{}$\\     
\hspace{3mm}\Nl{C}{iv}{1549}                            &1090,^{+348}_{-446}&944,\pm{3253}&0.87&666,\pm{2493}&0.61&$\cdot{}$\\  
\hspace{3mm}\Lya                                        &1790,^{+846}_{-831}&977,\pm{687}&0.55&1377,\pm{849}&0.77&$\cdot{}$\\   
\hmidline
\multicolumn{2}{l}{\hspace{-3mm}NGC 5548 -- 1988/1989 (opt+IUE)} &\multicolumn{4}{l}{($M_{\mathrm{BH}} = 67.1\times10^{6}\,  \mathrm{M}_{\odot}$, $R_{\mathrm{S}} = 7.65\times10^{-3}\,\mathrm{ld} = 19.82\times10^{12}\,\mathrm{cm}$)}\\
\hmidline
\hspace{3mm}\Nl{He}{ii}{1640}                           &496,^{+222}_{-235}&169,\pm{182}&0.34&196,\pm{182}&0.39&43.33$\,\pm{\,0.10}$\\    
\hspace{3mm}\Nl{He}{ii}{4686}                           &1019,^{+418}_{-392}&222,\pm{169}&0.22&209,\pm{169}&0.21&$\cdot{}$\\   
\hspace{3mm}\Nl{C}{iv}{1549}                            &1280,^{+248}_{-196}&1385,\pm{849}&1.08&1136,\pm{744}&0.89&$\cdot{}$\\ 
\hspace{3mm}\ion{Si}{iv}+\ion{O}{iv]}\,{$\lambda$1400}  &1607,^{+444}_{-392}&339,\pm{4600}&0.21&888,\pm{4639}&0.55&$\cdot{}$\\ 
\hspace{3mm}\Hb                                         &2574,^{+196}_{-196}&405,\pm{104}&0.16&431,\pm{104}&0.17&$\cdot{}$\\   
\hspace{3mm}\sVl{C}{iii}{1909}                          &3580,^{+705}_{-692}&1110,\pm{914}&0.31&1842,\pm{1097}&0.51&$\cdot{}$\\
 \hmidline
\multicolumn{3}{l}{\hspace{-3mm}NGC 5548 -- 1993 (opt+HST)}\\
 \hmidline
\hspace{3mm}\Nl{He}{ii}{1640}                           &248,^{+39}_{-39}&222,\pm{169}&0.89&117,\pm{130}&0.47&43.32$\,\pm{\,0.10}$\\      
\hspace{3mm}\ion{Si}{iv}+\ion{O}{iv]}\,{$\lambda$1400}  &561,^{+143}_{-130}&627,\pm{1097}&1.12&339,\pm{1019}&0.60&$\cdot{}$\\  
\hspace{3mm}\Nl{C}{iv}{1549}                            &875,^{+117}_{-130}&444,\pm{130}&0.51&653,\pm{143}&0.75&$\cdot{}$\\    
\hspace{3mm}\Hb                                         &1751,^{+496}_{-561}&431,\pm{196}&0.25&169,\pm{143}&0.10&$\cdot{}$\\   
\hspace{3mm}\sVl{C}{iii}{1909}                          &1816,^{+235}_{-182}&1816,\pm{1332}&1.00&1097,\pm{927}&0.60&$\cdot{}$\\
 \hmidline
\multicolumn{3}{l}{\hspace{-3mm}NGC 5548 -- H$\beta$}\\
 \hmidline
\hspace{3mm}\Hb                                         &849,^{+744}_{-483}&52,\pm{52}&0.06&91,\pm{91}&0.11     &43.05$\,\pm{\,0.11}$\\     
\hspace{3mm}\Hb                                         &1019,^{+496}_{-365}&222,\pm{222}&0.22&91,\pm{196}&0.09 &43.29$\,\pm{\,0.10}$\\ 
\hspace{3mm}\Hb                                         &1437,^{+248}_{-261}&261,\pm{78}&0.18&169,\pm{78}&0.12  &43.01$\,\pm{\,0.11}$\\  
\hspace{3mm}\Hb                                         &1751,^{+496}_{-561}&431,\pm{196}&0.25&169,\pm{143}&0.10&43.32$\,\pm{\,0.10}$\\
\hspace{3mm}\Hb                                         &1868,^{+771}_{-953}&182,\pm{143}&0.10&209,\pm{143}&0.11&43.05$\,\pm{\,0.11}$\\
\hspace{3mm}\Hb                                         &2077,^{+378}_{-326}&235,\pm{104}&0.11&248,\pm{104}&0.12&43.29$\,\pm{\,0.10}$\\
\hspace{3mm}\Hb                                         &2143,^{+156}_{-143}&222,\pm{104}&0.10&261,\pm{104}&0.12&43.37$\,\pm{\,0.09}$\\
\hspace{3mm}\Hb                                         &2286,^{+261}_{-209}&196,\pm{104}&0.09&287,\pm{104}&0.13&43.18$\,\pm{\,0.10}$\\
\hspace{3mm}\Hb                                         &2430,^{+274}_{-300}&392,\pm{143}&0.16&352,\pm{143}&0.15&43.08$\,\pm{\,0.11}$\\
\hspace{3mm}\Hb                                         &2574,^{+196}_{-196}&405,\pm{104}&0.16&431,\pm{104}&0.17&43.33$\,\pm{\,0.10}$\\
\hspace{3mm}\Hb                                         &2835,^{+339}_{-339}&378,\pm{196}&0.13&313,\pm{196}&0.11&43.46$\,\pm{\,0.09}$\\
\hspace{3mm}\Hb                                         &3240,^{+418}_{-392}&117,\pm{65}&0.04&352,\pm{78}&0.11  &43.44$\,\pm{\,0.09}$\\  
\hspace{3mm}\Hb                                         &3463,^{+561}_{-287}&392,\pm{235}&0.11&509,\pm{235}&0.15&43.52$\,\pm{\,0.09}$\\
\hbotline
\end{tabular}
\vspace{-0.5mm}
\end{table*}

As we did before for NGC~5548 (Paper III), we also calculate
corrected heights of the line-emitting
regions, which are
 based on the turbulent velocities belonging to  the individual lines
(see Paper II): 400 km\,s$^{-1}$ for H$\beta$,  
700 km\,s$^{-1}$ for H$\alpha$, 
900 km\,s$^{-1}$ for \ion{He}{ii}\,$\lambda 4686$,
2100 km\,s$^{-1}$ for  \ion{Si}{iv}+\ion{O}{iv]}\,$\lambda 1400$,
2300 km\,s$^{-1}$ for \ion{He}{ii}\,$\lambda 1640$,
2900 km\,s$^{-1}$ for \ion{C}{iv}\,$\lambda 1549$, and
3800 km\,s$^{-1}$ for Ly$\alpha$. 
The corrected heights $H_{\mathrm{corr}}$ of
 the line-emitting regions, which are based on these $v_{\mathrm{turb}}$,
are given in Table 1 as well.

The observed line widths FWHM of the individual emission lines
occur over a large range (Table~1, Figs.~1, 2) in the Seyfert galaxies
NGC~7469 and NGC~3783. 
The widths of their lines differ by a factor of 2.5 to 6.  
In contrast to the other galaxies,
the line widths of the emission lines in 3C~390.3
(Table~1, Fig.~3) differ by a factor of 1.18 only.
The radii of the line-emitting regions in 3C~390.3 (Table 1) also span a
comparatively limited range.
Another general trend has been noted before in other AGN that emit
 spectra with narrower line widths: namely, that the more highly ionized
lines are broader in comparison
to lower ionized lines and that they originate closer to the center
(e.g., Shuder\citealt{shuder82}, Gaskell \& Sparke\citealt{gaskell86},
Krolik\citealt{krolik91}, Kollatschny\citealt{kollatschny03}).
This trend is not to be seen in 3C~390.3.
%
%
\onecolumn
%
\begin{figure}
\begin{minipage}[t]{0.475\textwidth}
\includegraphics[width=5.5cm,angle=0]{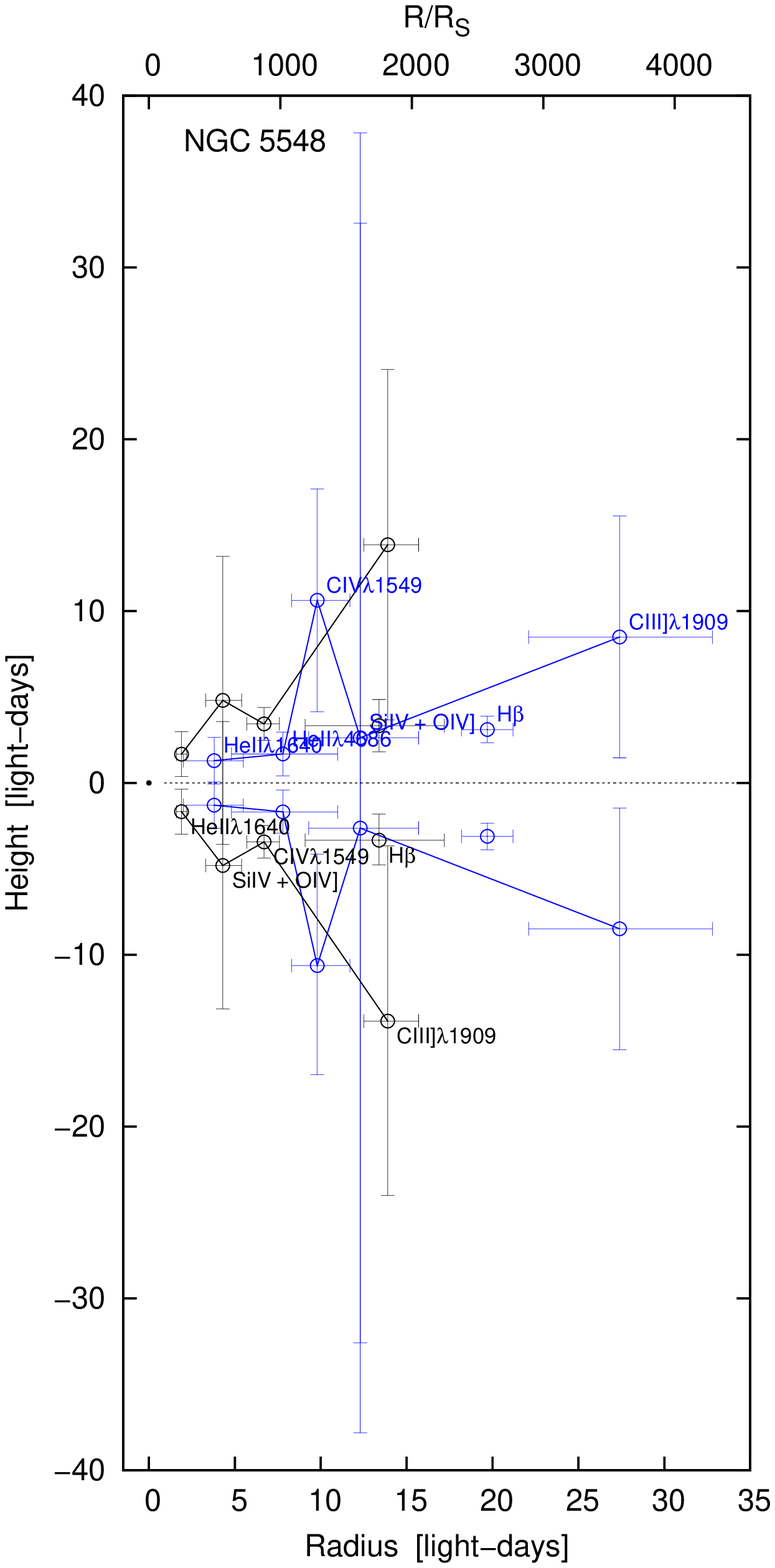} 
  \caption{NGC~5548 broad-line region structure as a function of distance
 to the center as well as height above the midplane for the two 
   variability campaigns in 1988/89 (blue) and 1993 (black).
 The highly ionized lines are connected by a solid line.
The Balmer lines are kept separately.
The dot at radius zero gives the size of a Schwarzschild black hole (with
$M=6.71\times10^{7}M_{\sun}$) multiplied by a factor of twenty.}
   \label{disc_ngc5548_t2.eps}
\end{minipage}
\hfill
\begin{minipage}[t]{0.475\textwidth}
\includegraphics[width=5.5cm,angle=0]{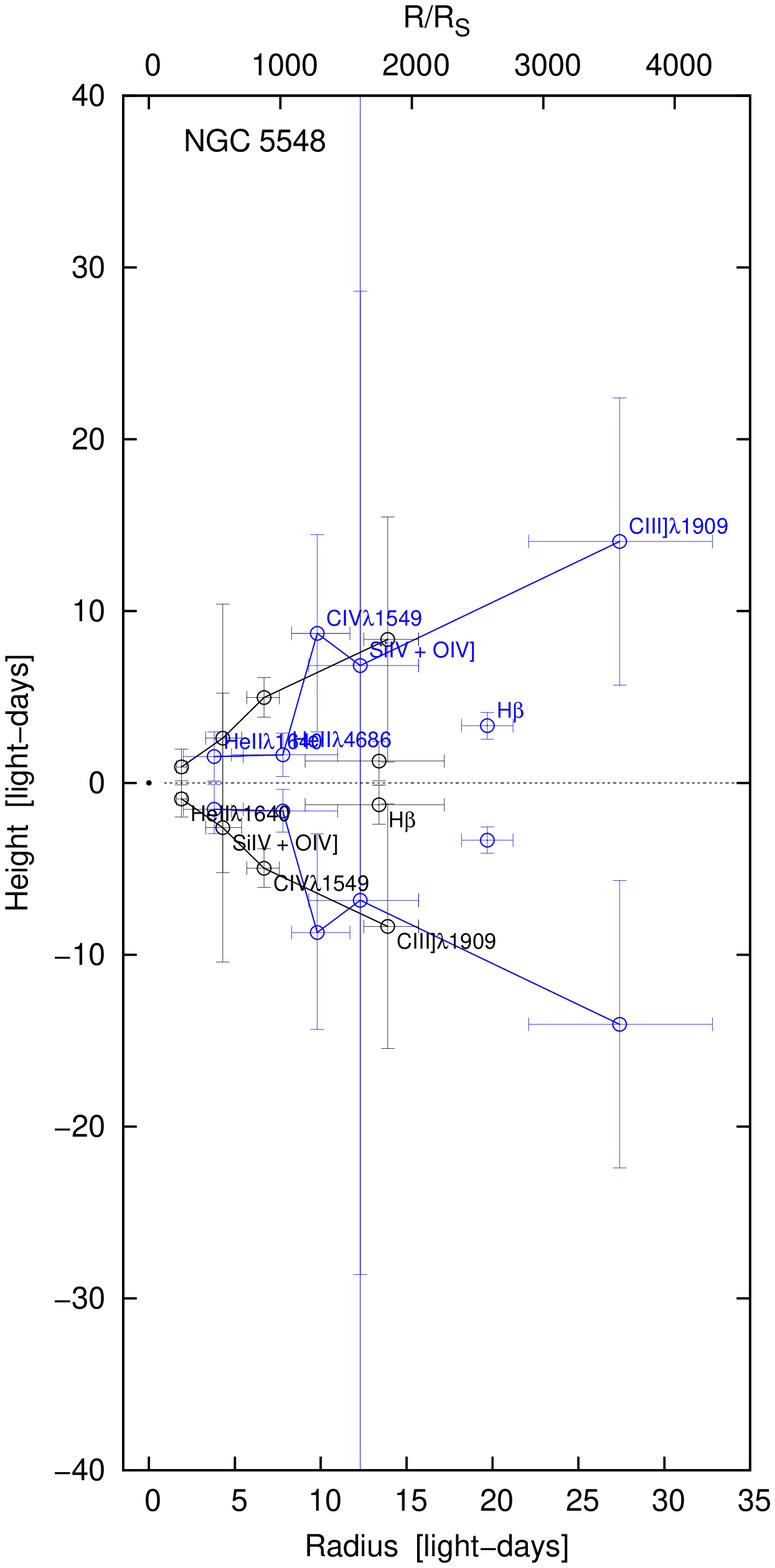} 
  \caption{NGC~5548 broad-line region structure - for the two 
   variability campaigns in 1988/89 (blue) and 1993 (black).
 Same as Fig. 4, but based on corrected turbulent velocities $v_{\mathrm{turb}}$.}
   \label{disc_ngc5548_t2corr.eps}
\end{minipage}
%
%
%
%
%
\begin{minipage}[t]{0.475\textwidth}
\includegraphics[width=5.5cm,angle=0]{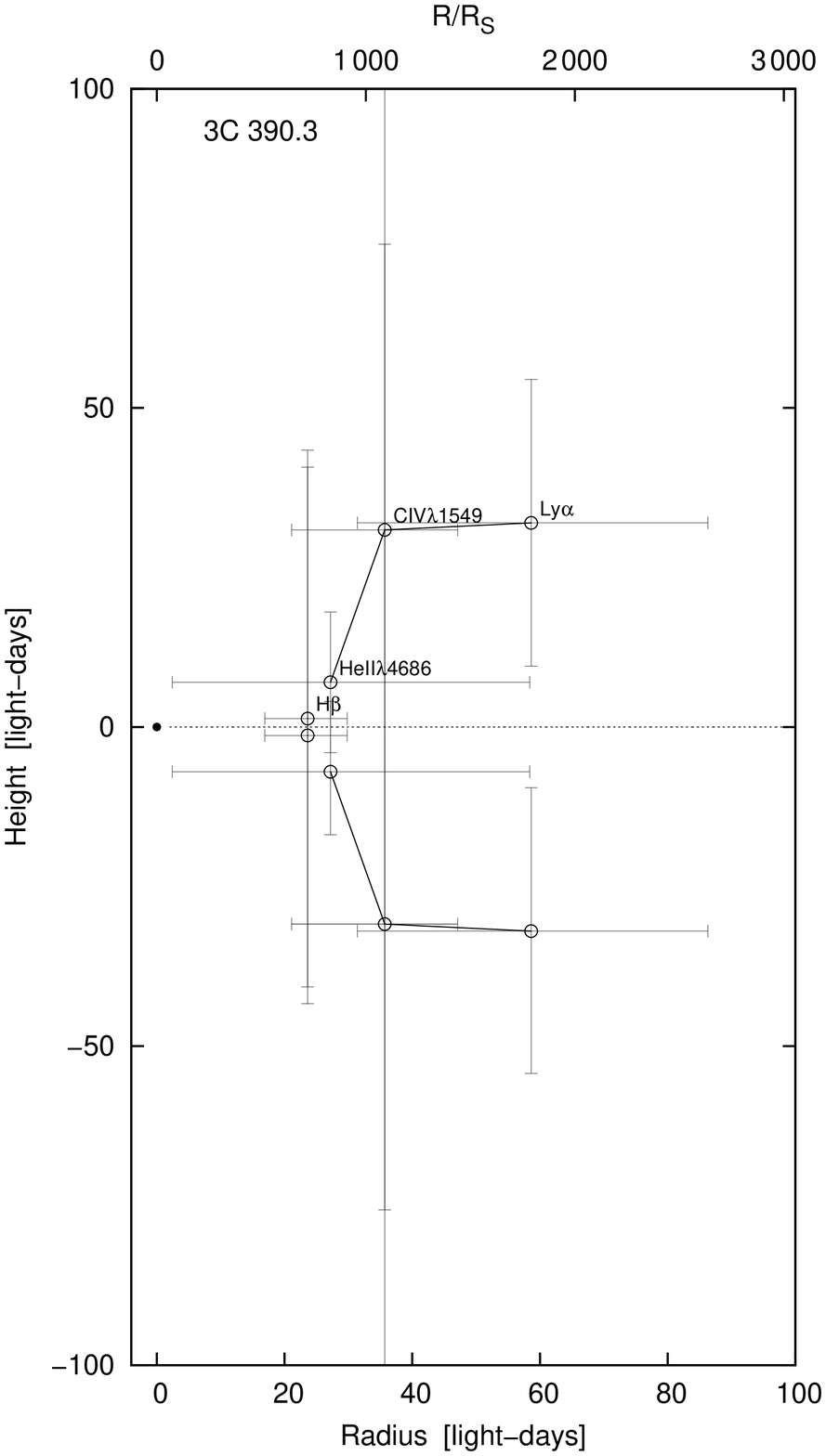} 
  \caption{3C~390.3 broad-line region structure.
 The highly ionized lines are connected by a solid line.
 H$\beta$ has been kept separately.
The dot at radius zero has the size of a Schwarzschild black hole (with
$M=28.7\times10^{7}M_{\sun}$) multiplied by a factor of twenty.}
   \label{disc_3c390.eps}
\vspace{5mm} 
\end{minipage}
\hfill
\begin{minipage}[t]{0.475\textwidth}
\includegraphics[width=5.5cm,angle=0]{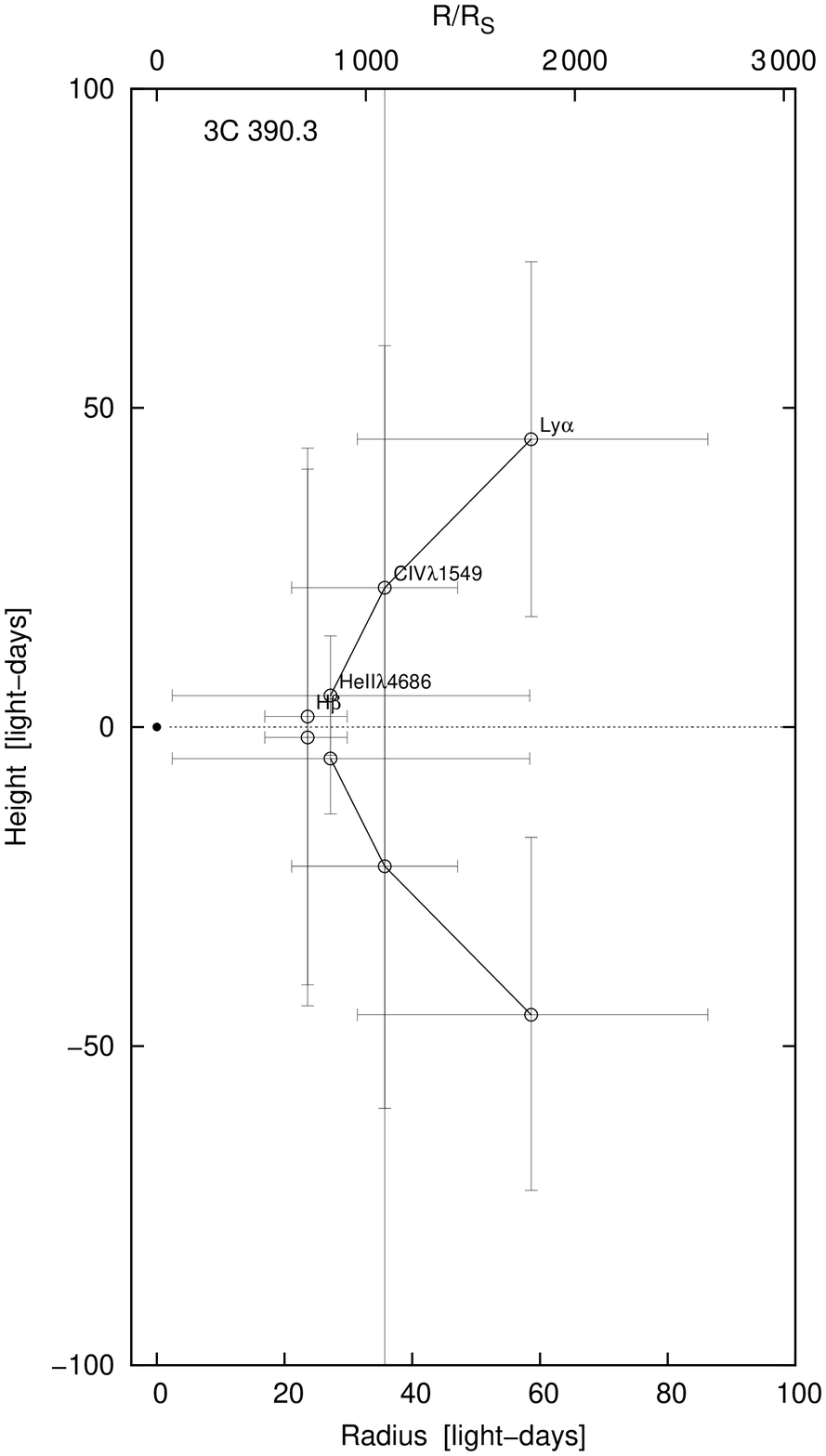} 
  \caption{3C~390.3 broad-line region structure. Same as Fig. 6, but
   based on corrected turbulent velocities $v_{\mathrm{turb}}$.} 
   \label{disc_3c390corr.eps}
\end{minipage}
\end{figure}
%
%

%
\begin{figure}
\begin{minipage}[t]{0.475\textwidth}
\includegraphics[width=7.6cm,angle=0]{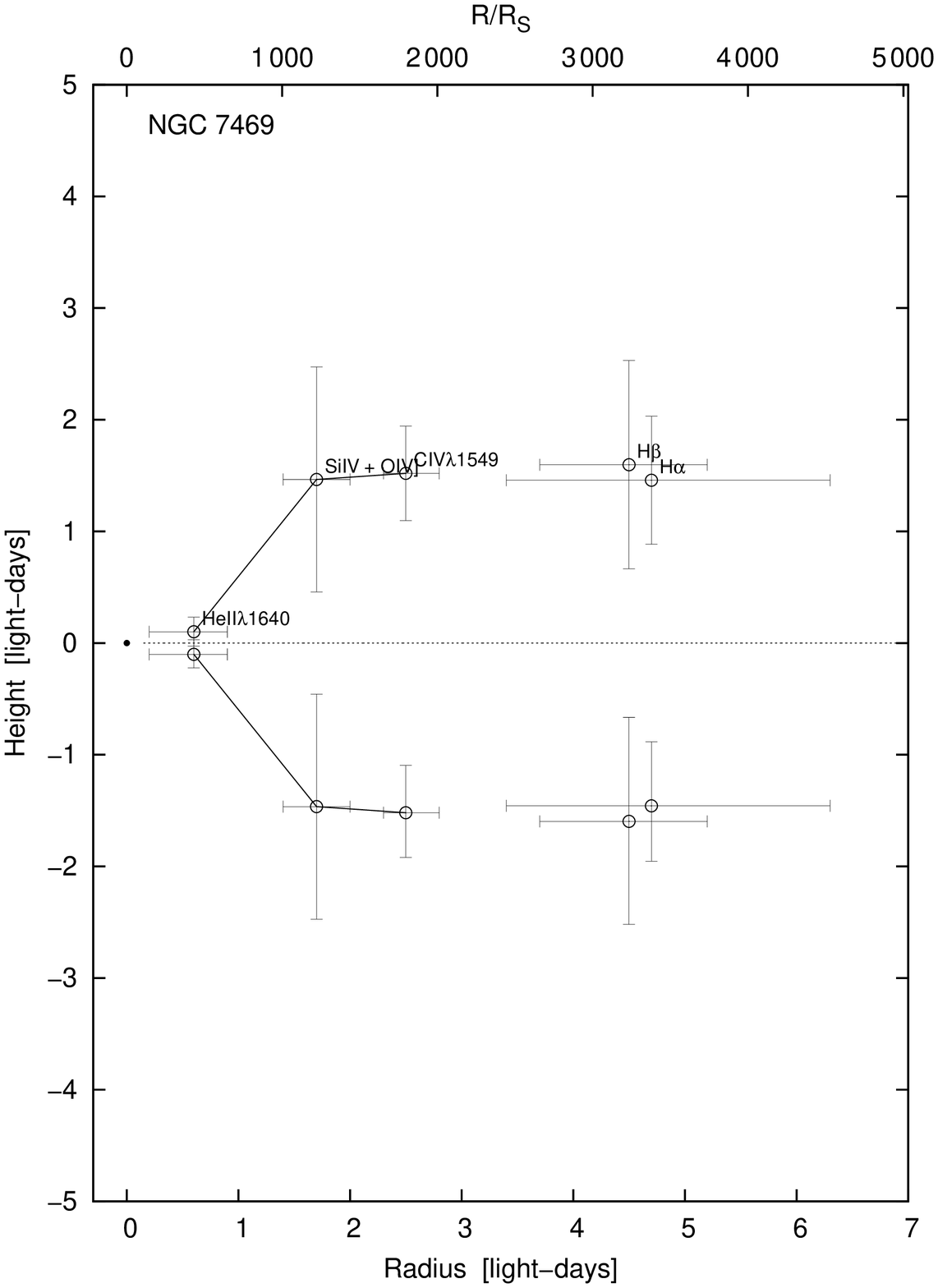} 
  \caption{NGC~7469 broad-line region structure as a function of distance
 to the center as well as height above the midplane. The highly ionized lines
 are connected by a solid line. The Balmer lines are kept separately.
The dot at radius zero has the size of a Schwarzschild black hole (with
$M=1.22\times10^{7}M_{\sun}$) multiplied by a factor of twenty.}
   \label{disc_ngc7469.eps}
\vspace{5mm} 
\end{minipage}
\hfill
\begin{minipage}[t]{0.475\textwidth}
\includegraphics[width=7.6cm,angle=0]{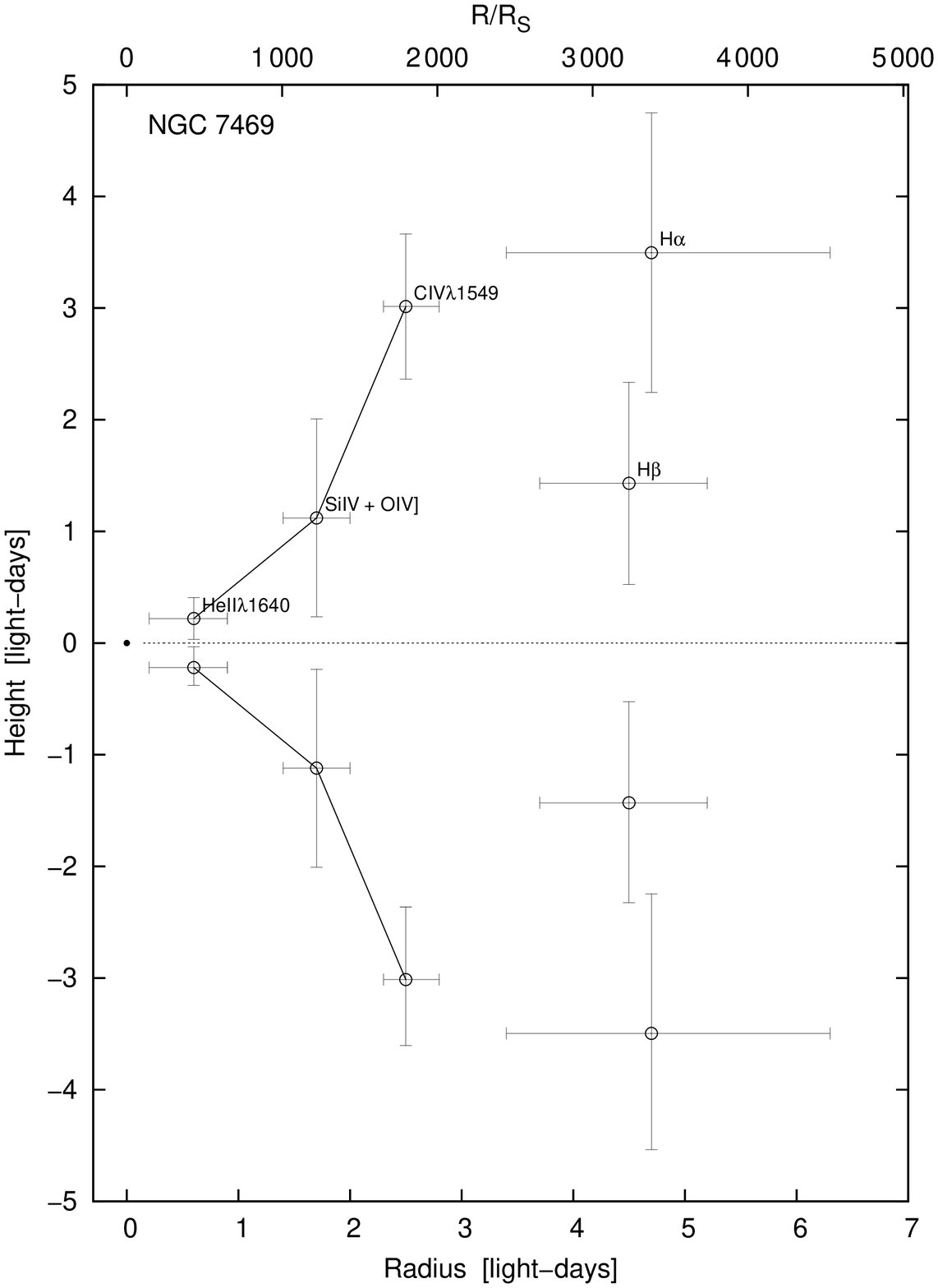} 
  \caption{NGC~7469 broad-line region structure. Same as Fig. 8, but
   based on corrected turbulent velocities $v_{\mathrm{turb}}$.} 
   \label{disc_ngc7469corr.eps}
\end{minipage}
%
%
%
\begin{minipage}[t]{0.475\textwidth}
\includegraphics[width=7.6cm,angle=0]{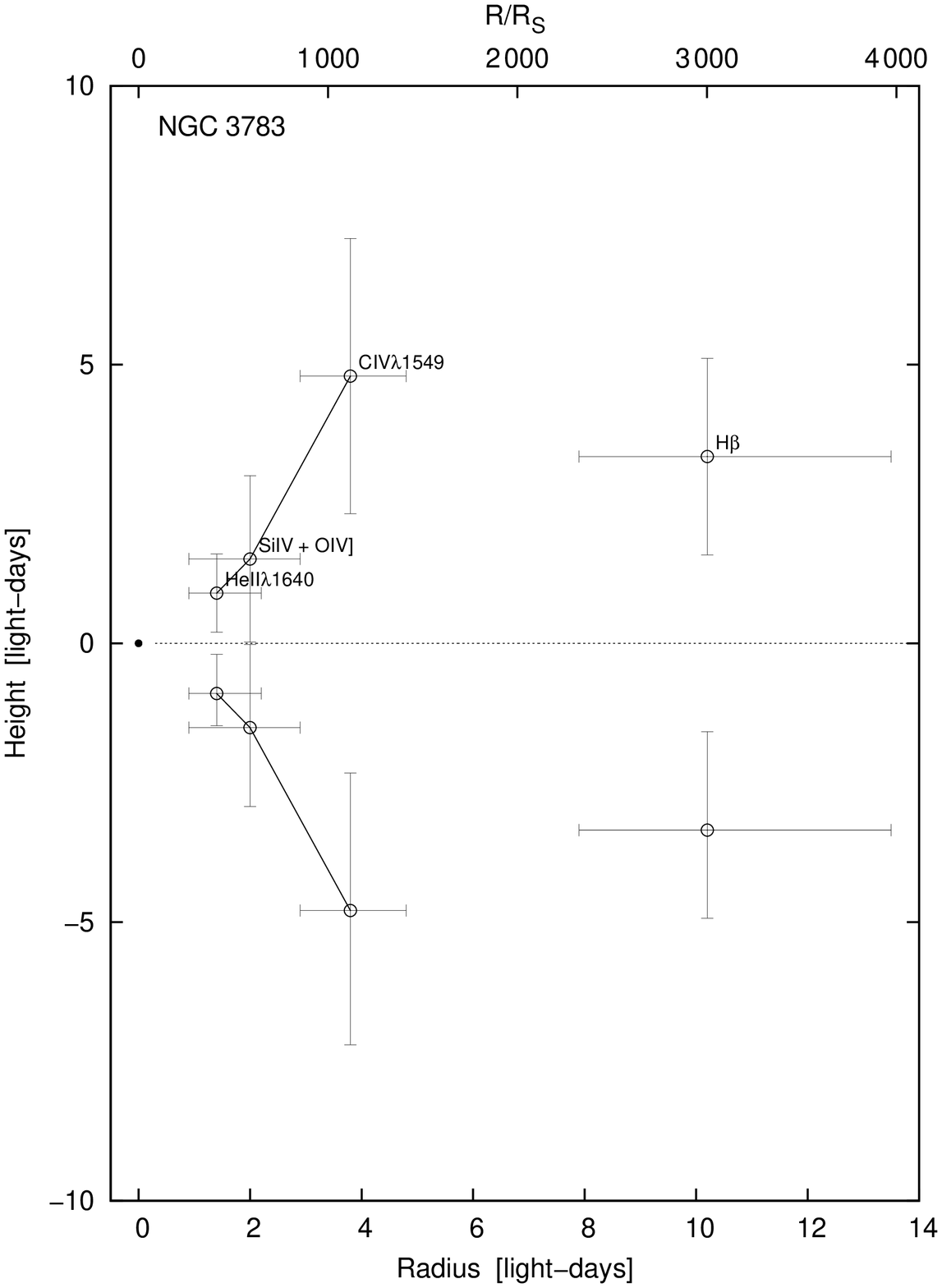} 
  \caption{NGC~3783 broad-line region structure.
   The highly ionized lines are connected by a solid line.
    H$\beta$ has been kept separately.
The dot at radius zero has the size of a Schwarzschild black hole (with
$M=2.98\times10^{7}M_{\sun}$) multiplied by a factor of twenty.}
   \label{disc_ngc3783.eps}
\end{minipage}
\hfill
\begin{minipage}[t]{0.475\textwidth}
\includegraphics[width=7.6cm,angle=0]{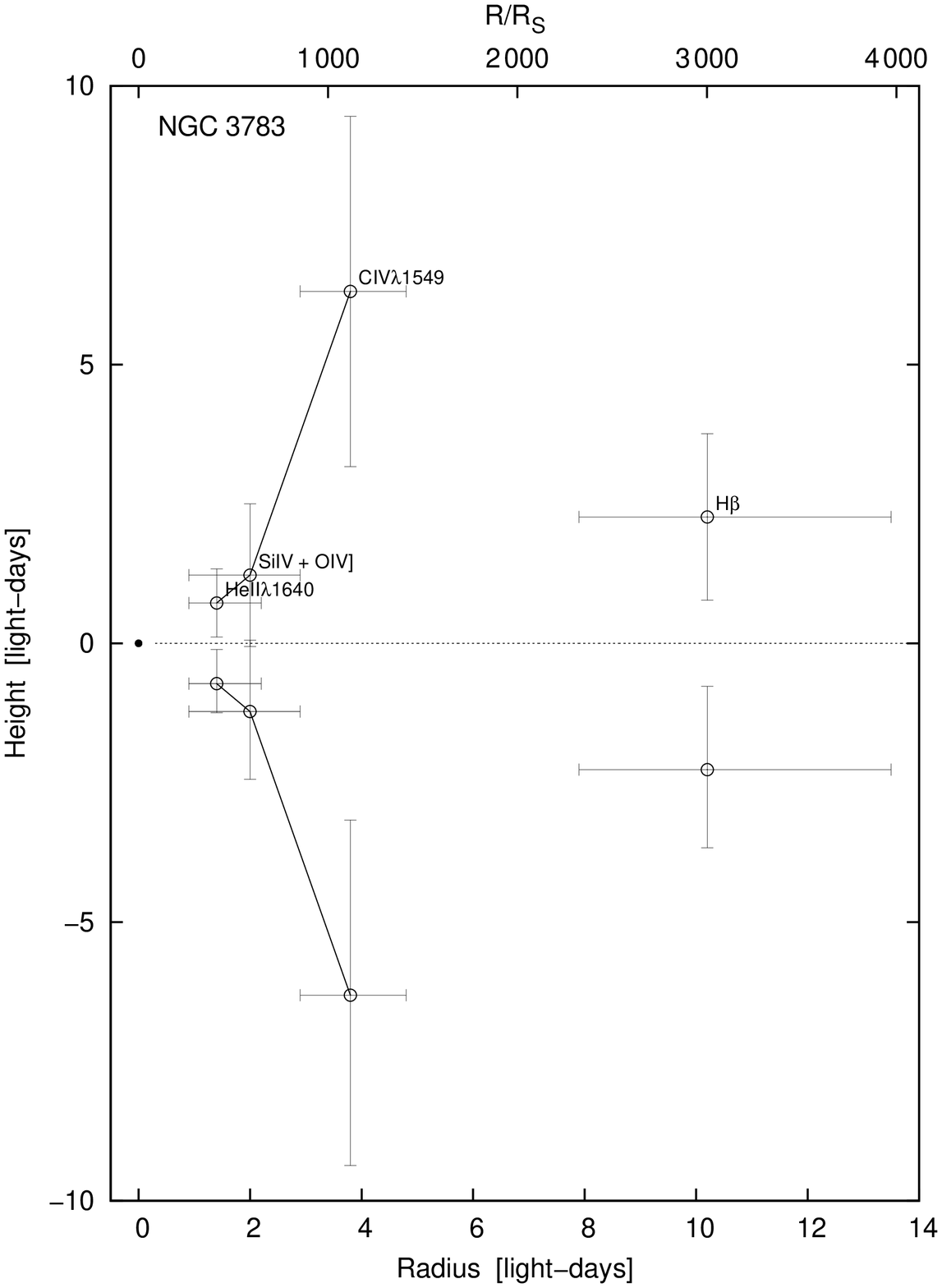} 
  \caption{NGC~3783 broad-line region structure. Same as Fig. 10, but
   based on corrected turbulent velocities $v_{\mathrm{turb}}$.} 
   \label{disc_ngc3783corr.eps}
\end{minipage}
\end{figure}
\twocolumn
%
%

\subsection{Broad-line region geometries in
NGC~7469, NGC~3783, and 3C~390.3 in
 comparison to NGC~5548}

The broad-line region geometry in NGC~5548, which is based on two independent
optical/UV variability campaigns, has been presented and discussed
in Paper III. 
Different emission lines originate in different regions and at different
distances from the center in NGC~5548. The distances of the emitting regions
depend, among other things, on the luminosity of the central ionizing source.
We connected the emitting regions of the individual emission lines
by solid lines to demonstrate this effect (see Fig.~5 in Paper III).
Furthermore, we could verify the general trend
 that more highly ionized lines
originate in a more extended shell above the accretion disk, while Balmer
lines originate in an independent region far closer to the accretion disk.
Therefore it appears to be justified to connect the emitting line regions
of the highly ionized lines  
to get an illustration of the broad-line region structure.
We present the broad-line region structure of NGC~5548 in Figs. 4 and 5
so that we connect the line-emitting regions of the highly
ionized lines separately for the two variability campaigns. 
The Balmer line-emitting regions are kept detached.
The line-emitting regions based on the 
 variability campaign in 1988/89  are given in blue and those of the
 variability campaign in 1993 are given in black.
 The highly ionized lines (i.e., the non-Balmer lines) are connected
 by a solid line.
The dot at radius zero gives the size of a Schwarzschild black hole (with
$ M=6.7\times10^{7}M_{\odot}$ taken from Peterson et al.\citealt{peterson04})
multiplied by a factor of twenty.
We make the assumption that the accretion disk structures are arranged
symmetrically to the midplane. 
The two axes scale  in Figs.~4 and 5 are linear in units of light-days.
One light-day corresponds to a distance of $2.59\times{}10^{15}$\,cm.
The axis on top of the figure gives the distance of the line-emitting regions
from the center in units of the
Schwarzschild radius.
The left-hand diagrams (Figs.~4, 6, 8, 10) are given for the
 observed turbulent
velocities $v_{\mathrm{turb}}$, while the right-hand diagrams (Figs.~5,
7, 9, 11)
are shown for the average of the turbulent velocities belonging to the
individual emitting line regions (see Papers I to III).

Optical/UV spectra were taken during one variability campaign only
for the Seyfert galaxies NGC~7469 and NGC~3783, as well as for 3C~390.3. 
Insofar as we cannot connect emitting regions of individual lines
we now combine the sites of all highly ionized lines in 3C~390.3,
NGC~7469, and NGC~3783, similar to what we did
in NGC~5548 (Figs. 4 and 5).
These diagrams (Figs. 6 to 11) present their individual 
broad-line region structures as a function of distance
 to the center as well as height above the midplane.
The line-emitting regions of the Balmer lines have been kept separately.
The dots at radius zero give the corresponding sizes of their
Schwarzschild black holes (taken from Peterson et al.\citealt{peterson04})
 multiplied by a factor of twenty.
The numerical values of the Schwarzschild radii as well as 
line-emitting regions are shown in Tables 1 and 2.
Furthermore, the optical continuum luminosities at 5100 \AA{}
(corrected for the contribution of the host galaxies:
Bentz et al.\citealt{bentz13}) are given
for the individual variability campaigns of the AGN.

We present in Fig. 4 to 11 the broad emitting line geometries
for observed as well as for
corrected turbulent velocities (see Papers II, III). There are
small differences in their particular geometries. 
However, the general trends in their structures are identical.


\section{Discussion}


The derived BLR structures of the three AGN NGC~7469, NGC~3783, and 3C~390.3
 (Figs. 6 to 11) confirm
trends in their emitting region geometry
that have been noticed in NGC~5548 before (Paper III, Figs. 4 and 5):
The more highly ionized lines originate closer to
the central ionizing source and/or at larger distances above the midplane.
The Balmer lines are emitted closer to the midplane in a more flattened
configuration in comparison to the other emission lines.
It should be emphasized that the individual emission lines do not originate
at one single radius only but rather in an extended region, depending
on the luminosity of the central ionizing source 
(see, e.g.,  Kollatschny\citealt{kollatschny03} and Paper III).
However, besides these common trends in the geometries
of the line-emitting regions, there are differences  
with respect to the individual galaxies.

\subsection{Comparison of the broad-line region
 geometries in different Seyfert galaxies}

We jointly present the broad-line region structures
of NGC~7469, NGC~3783,  NGC~5548 (two epochs), and 3C~390.3
(in units of light-days) in a single diagram in Figs. 12 and 13 for
comparison of
their geometries.
The two axes scale are linear in units of light-days.
The emitting regions of the
 highly ionized lines in the individual galaxies are connected
 by solid lines, as done in Figures 4 to 11.
The H$\beta$ emitting line regions have been kept separately. We 
connect the H$\beta$ emitting regions
of NGC~5548 for the 13 observing epochs as done in Paper III.
The H$\beta$ emitting regions are drawn in red.

Besides the fact that the emission line regions of the galaxies show
different extensions in radius and originate at different heights
above the midplane, there is
a further systematic trend to be seen in these diagrams.
 This trend goes with the line widths FWHM (e.g., H$\beta$ and/or
\ion{C}{iv}\,$\lambda 1549$) of the galaxies.  
The `narrow' line Seyfert galaxies NGC~7469 and NGC~3783 show 
H$\beta$ line widths (FWHM) of 2170 to 3100 km\,s$^{-1}$ and 
\ion{C}{iv}\,$\lambda 1549$ line widths of 3690 to 4300 km\,s$^{-1}$.
 NGC~5548 has 
H$\beta$ line widths of 4040 to 8050 km\,s$^{-1}$ and 
\ion{C}{iv}\,$\lambda 1549$ line widths of 6560 to 6870 km\,s$^{-1}$
(Paper III).
3C~390.3 exhibits a H$\beta$ line width of 9960 km\,s$^{-1}$ and a 
\ion{C}{iv}\,$\lambda 1549$ line width of 8990 km\,s$^{-1}$.
The narrow line Seyferts NGC~7469 and NGC~3783 show narrower cone opening
angles along the rotation axis
concerning the highly ionized emitting line regions
(in Figs. 12, 13)
in comparison to the broad-line Seyferts NGC~5548 and 3C~390.3
 As the line widths are related to the rotational velocities of the
line-emitting regions, this sequence is one
of increasing central rotation.
Furthermore, the high-ionization lines of the broad-line Seyferts
NGC~5548 and 3C~390.3 originate at larger radii further outside.

The H$\beta$ emitting regions always occur at lower height-to-radius ratios
(up to a factor of ten)
and originate closer to the midplane than the high-ionization lines do.
Based on theoretical models, Netzer\cite{netzer90} has discussed
that the high-ionization broad emission lines and most of the 
Ly$\alpha$ flux come from a spherical system 
above the midplane in comparison
to the low-ionization lines (his Fig.~13).
The height-to-radius ratio for H$\beta$ is smallest
for 3C~390.3 (with 0.07) and largest
for the narrow line Seyferts NGC~7469 and NGC~3783 
(0.33 and 0.50).
%
%
\vspace*{5mm} 
\begin{figure}
\includegraphics[width=8.0cm,angle=0]{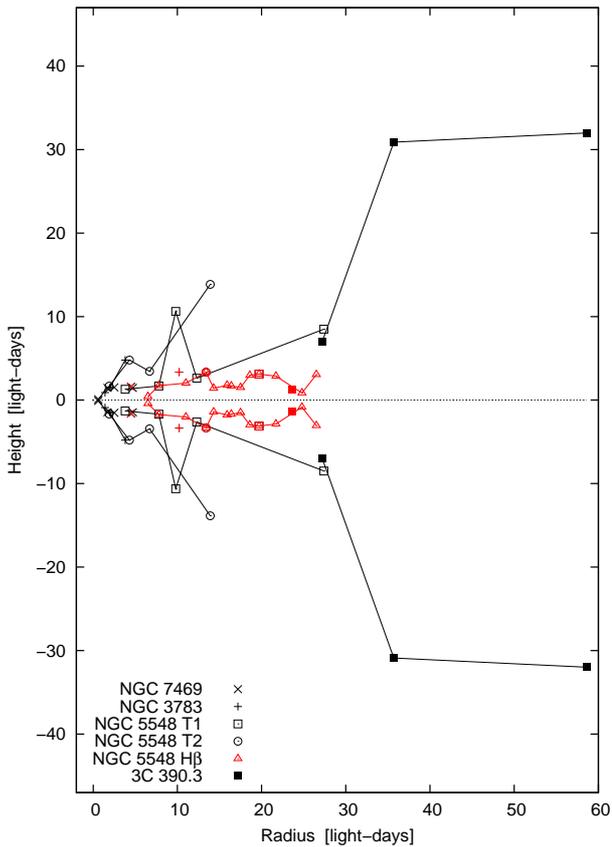} 
  \caption{Broad-line region structures in NGC~7469, NGC~3783, 
NGC~5548 (two epochs), and 3C~390.3
as a function of distance to the center as well as height above the midplane.
The highly ionized (non-Balmer) lines of the individual galaxies are connected
 by a solid line. The H$\beta$ emitting regions are drawn in red.
 The H$\beta$ emitting line regions
of NGC~5548 (13 epochs) are connected by a solid red line.
}
   \label{disc_all1.eps}
\end{figure}
%
%
\vspace*{5mm} 
\begin{figure}
\includegraphics[width=8.0cm,angle=0]{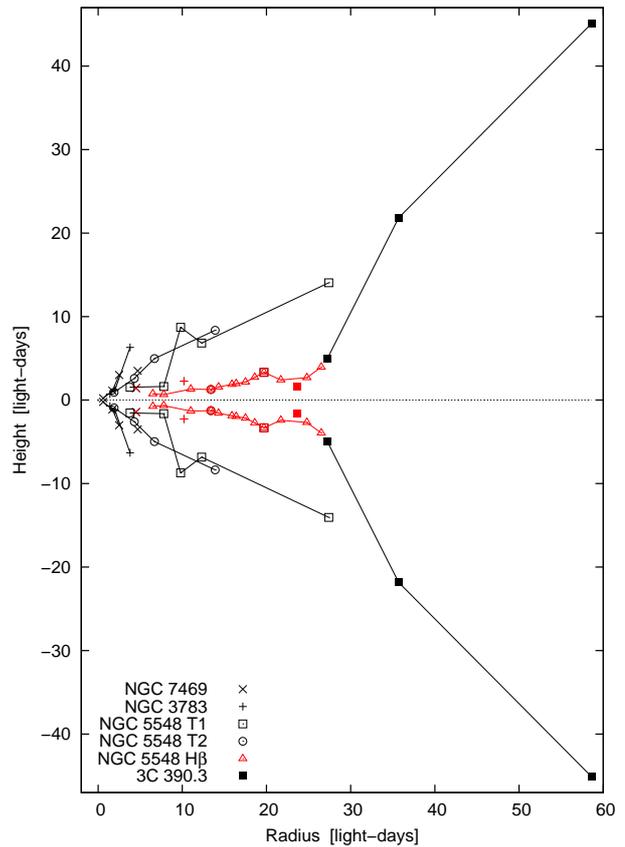} 
  \caption{Broad-line region structures in NGC~7469, NGC~3783, 
NGC~5548 (two epochs), and 3C~390.3.
Same as Fig. 12, but based on corrected turbulent velocities $v_{\mathrm{turb}}$.}
   \label{disc_all1corr.eps}
\end{figure}
%

 All high-ionization lines originate at larger
radii in 3C~390.3 compared to the rest of the galaxies.    
Furthermore, the broad-line region structure of 3C~390.3 is different
in one aspect
in comparison to the other AGN: the H$\beta$ line shows a smaller lag
and broader Doppler width than the \ion{C}{iv}\,$\lambda 1549$  line. This
has been noticed before by  Peterson \& Wandel\cite{peterson00}.
The reported delay of 23.6 days (Dietrich et al.\citealt{dietrich98})
of the broad H$\beta$ emission flux with respect to the optical continuum
based on optical data for the years 1994/1995 was short-term
in comparison to other
variability campaigns of this galaxy.
Other long-term spectral monitoring campaigns of this galaxy for the years
1995-2007 (Shapovalova et al.\citealt{shapovalova10},
Popovic et al.\citealt{popovic11})
and 1992-2000 (Sergeev et al.\citealt{sergeev02}) resulted in delays of about
90 days of the broad H$\beta$ emission flux with respect to the optical
continuum.
The size relationship of the broad-line region structure in 3C~390.3
(Figs. 6,7) would be very similar to that of the other galaxies (Figs. 4, 5,
and 8 -- 11)
if the H$\beta$ size were of the order of 80 light-days instead of 24 days.
 This size value would be
more consistent with the results of Popovic et al.\cite{popovic11}.
On the other hand, the derived radius of 24 light-days is consistent 
with the observed luminosity in the BLR size–luminosity relation (see Fig.~16).
In a recent investigation, Dietrich et al.\cite{dietrich12} discussed the
results of another variability campaign of 3C~390.3 in 2005.
They determined a delay of 44 days for this campaign.
The continuum luminosity was about 6x stronger than in 2005 compared to 1995,
which is again consistent with the radius-luminosity relation of AGN.
Dietrich et al.\cite{dietrich12} demanded a dedicated long-duration campaign
of 3C~390.3 with densely sampled measurements to make firm statements
about the H$\beta$ size.
%
%
%
To investigate whether 3C~390.3 is an exception
with respect to the optical/UV line delays,
it is necessary to carry out simultaneous
\ion{C}{iv}\,$\lambda 1549$ and
H$\beta$ line variability studies
in further broad-line/double-peaked AGN as well.

The overall picture we derived for the broad-line region structure
before in NGC~5548
(Paper III) is strengthened by the emission line data of the galaxies NGC~7469,
NGC~3783, 
and 3C~390.3. The H$\beta$ line is emitted in a more flattened configuration
above the midplane in comparison to the highly ionized lines. 
The H$\beta$ lines originate at heights $H$
of 0.7 to 1.6 light-days only and at distances $R$ of 1.4 to 24 light-days
from the center (see Figs. 12, 13). This corresponds
to $H/R$ values of 0.07 to
0.5 (Table 1) for the H$\beta$ line-emitting regions.
These values are upper limits of the geometrical heights of the
associated accretion disks as they originate above the accretion disk.
 Additional
contributions of outflowing wind components to the line widths would reduce
the contribution of the turbulent velocities to the line profiles and
thus reduce the numerical values of the geometrical heights.

The highly ionized lines originate at smaller radii than the H$\beta$ lines
and/or at greater distances from the midplane at $H/R$ values of 0.2 to
1.7 (Table 1). Again, the same picture is confirmed
 (as seen before in NGC~5548) 
that the emission lines do not originate in a thin
atmosphere of an accretion disk but rather in very extended regions
above an accretion disk.
This is in accordance with the BLR model presented
 by Gaskell\cite{gaskell09}.
The observed geometries of the line-emitting regions resemble
the geometries of accretion disk wind models
(e.g., Murray \& Chiang\citealt{murray97}, Proga \& Kallman\citealt{proga04}).
There is a second trend to be seen when comparing the geometries of the
different galaxies: the angle of the central opening cone (generated by the
emitting regions of the highly ionized lines)
is smaller for the galaxies with slow rotational velocities
(based on their H$\beta$ and/or
\ion{C}{iv}\,$\lambda 1549$ line widths)
 and increases
with the rotation. 

\subsection{Comparison of broad-line region
 geometries in different Seyfert galaxies scaled to
their Schwarzschild radii}

%
%
%
%
\begin{figure}
\includegraphics[width=8.0cm,angle=0]{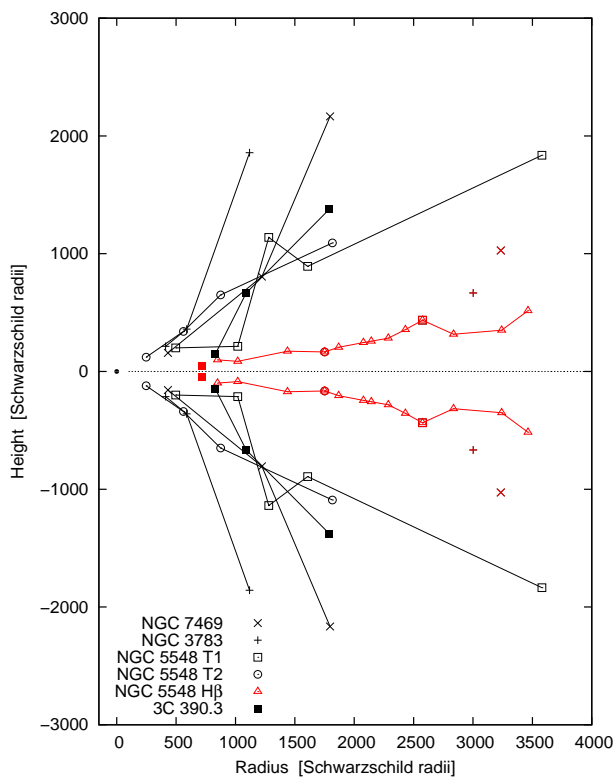} 
  \caption{Broad-line region structures in NGC~7469, NGC~3783, 
NGC~5548 (two epochs), and 3C~390.3
 scaled with respect to their individual
Schwarzschild black hole radii (based on corrected turbulent velocities
 $v_{\mathrm{turb}}$).
The emitting regions of the highly ionized lines of the individual galaxies are
 connected by a solid line.
 The H$\beta$ emitting regions are drawn in red. Those
 of  NGC~5548 are connected by a solid red line.
 The dot at radius zero gives the size of a black hole 
  with 20 Schwarzschild radii.}
   \label{disc_all2corr.eps}
\end{figure}
\begin{figure}
\includegraphics[width=8.0cm,angle=0]{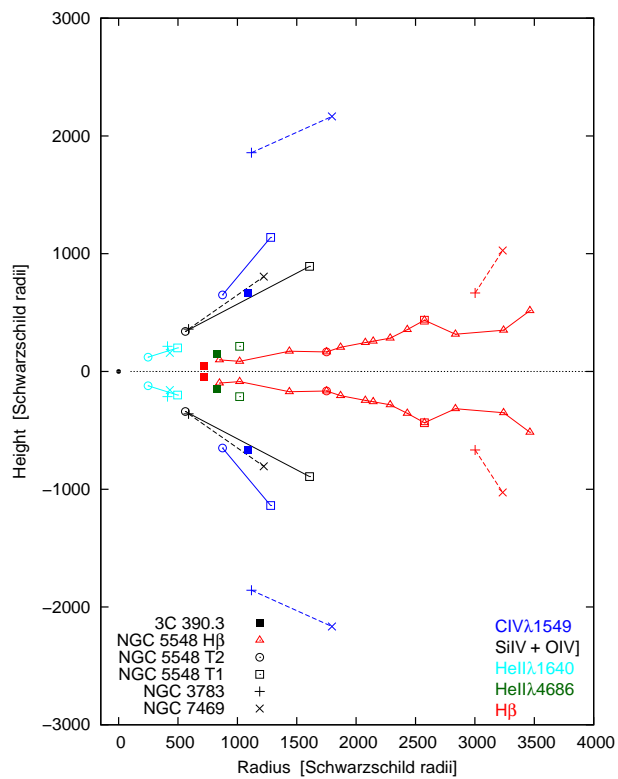} 
  \caption{Broad-line region structures in NGC~7469, NGC~3783, 
NGC~5548 (two epochs), and 3C~390.3
scaled to their individual Schwarzschild black hole radii.
Same as Fig.~14, but the emitting regions of the different lines are
drawn in different colors: \ion{C}{iv}\,$\lambda 1549$ (blue), 
\ion{Si}{iv}+\ion{O}{iv]}\,$\lambda 1400$ (black), 
\ion{He}{ii}\,$\lambda 1640$ (cyan),
\ion{He}{ii}\,$\lambda 4686$ (green),
H$\beta$ (red).
The individual line-emitting regions of the two narrow
line Seyfert galaxies NGC~7469 and NGC~3783 are connected by a dashed
line to aid the eye. 
The line-emitting regions
 of NGC~5548 are connected by a solid line.}
   \label{disc_all_same-ion3_sr.ps}
\end{figure}
%

Now we investigate the question whether BLR structures in AGN
are unambiguously characterized by their 
central black hole masses and by their central luminosities,
which control the distances/radii of the line-emitting regions. 
We jointly present in  Figs.~14 and 15  the BLR structures of our four AGN
scaled to their individual Schwarzschild masses and therefore to the same
black hole radii.
Again these BLR structures are based on the observed high-ionization
and H$\beta$ emission lines.
We use those black hole masses for calculating their
Schwarzschild radii 
that were derived by Peterson\cite{peterson04}.
These masses are still uncorrected regarding the contribution of the
turbulent velocities to their line widths.
 Furthermore,
the distances presented in Table 2 and Figures 14 to 15
were computed under the assumption 
that the geometrical scale factor $f$ for
calculating the central black hole masses $M_{\mathrm{BH}}$
(Peterson et al.\citealt{peterson04})
and their Schwarzschild radii is identical for all our galaxies:
\begin{equation}
\label{eq:MBH}
M_{\mathrm{BH}}  = f\,c\,\tau_{\mathrm{cent}}\,\Delta\,v^{2}\, G^{-1} ,
\end{equation}
with  $\tau_{cent}$ 
the characteristic distance of the line-emitting region,
$\Delta\,v$ the emission line width, $c$ the speed of
light, and $G$ the gravitational constant.\\
However, the scale factor $f$ in the formula depends on the
structure, kinematics, and orientation of the broad-line region and is
therefore different for all individual galaxies.
For a discussion, see Peterson et al.\cite{peterson04},
Kollatschny\cite{kollatschny03}, Collin et al.\cite{collin06},
Decarli et al.\cite{decarli08},
Goad et al.\cite{goad12} and references therein.
Even for the average factor  $<f>$,
diverse values are discussed ranging from $f=1$ 
(McLure \& Dunlop\citealt{mclure04}) to $f\sim5.5$
(Onken et al.\citealt{onken04}).
The relative black hole masses and therefore their radii are incorrect
on the order of a factor of a few in Figs. 14 to 15.

The extensions of the broad emission line regions
among themselves in all four AGN related to their 
Schwarzschild radii (Figs.~14, 15)
 are not that different in comparison to the
real extensions in units of light-days (Figs.~12, 13).
However, the general trends regarding the heights of the emitting regions
above the midplane remain the same as seen before.

Fig.~15 again shows a comparison of the BLR structures of our galaxies
(based on their strongest emission lines and on their line widths)
as a function of their central Schwarzschild radius.
%
%
\begin{figure*}
\hbox{
\includegraphics[width=6.0cm,angle=-90]{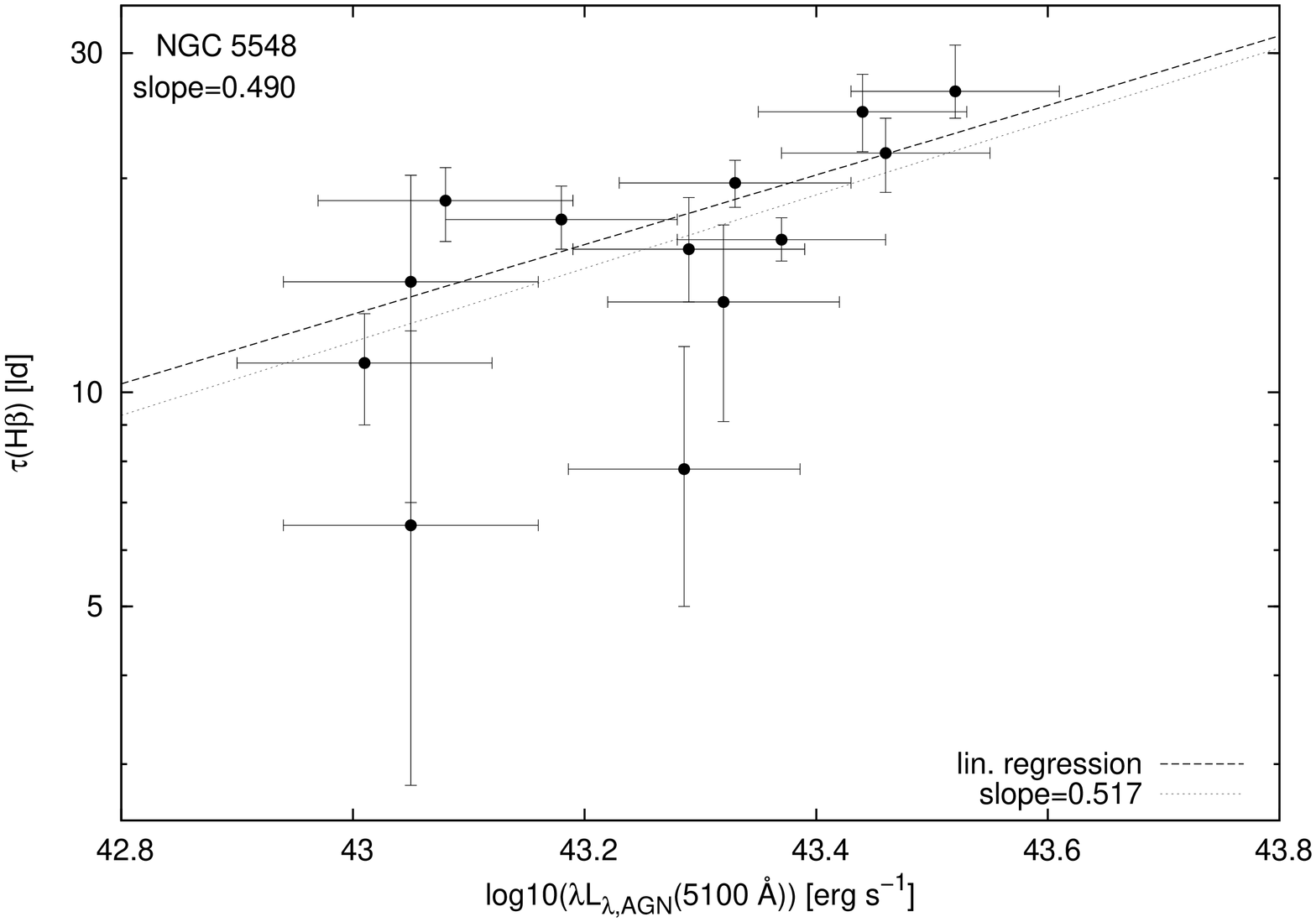} 
\includegraphics[width=6.0cm,angle=-90]{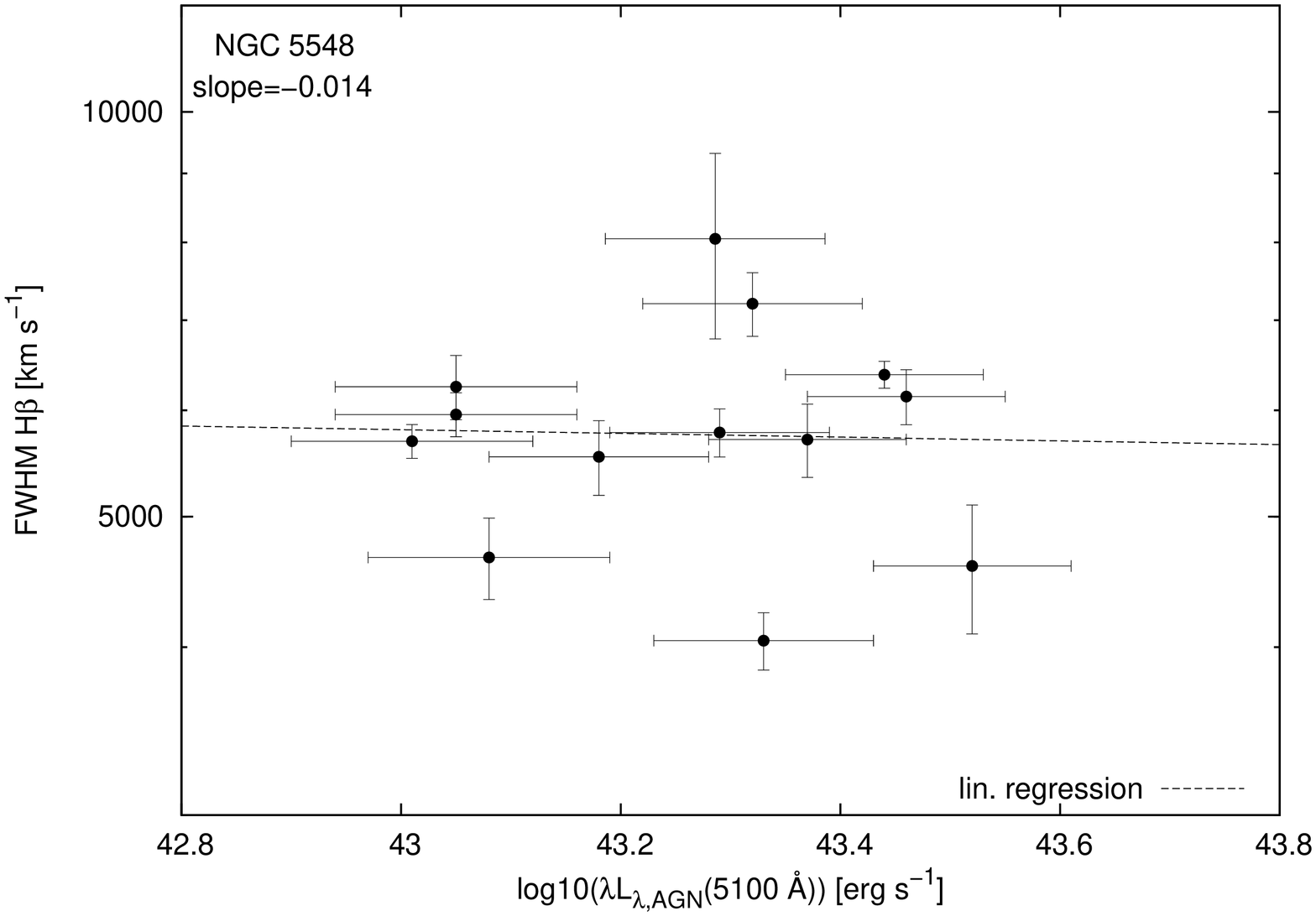} 
}\hbox{
\includegraphics[width=6.0cm,angle=-90]{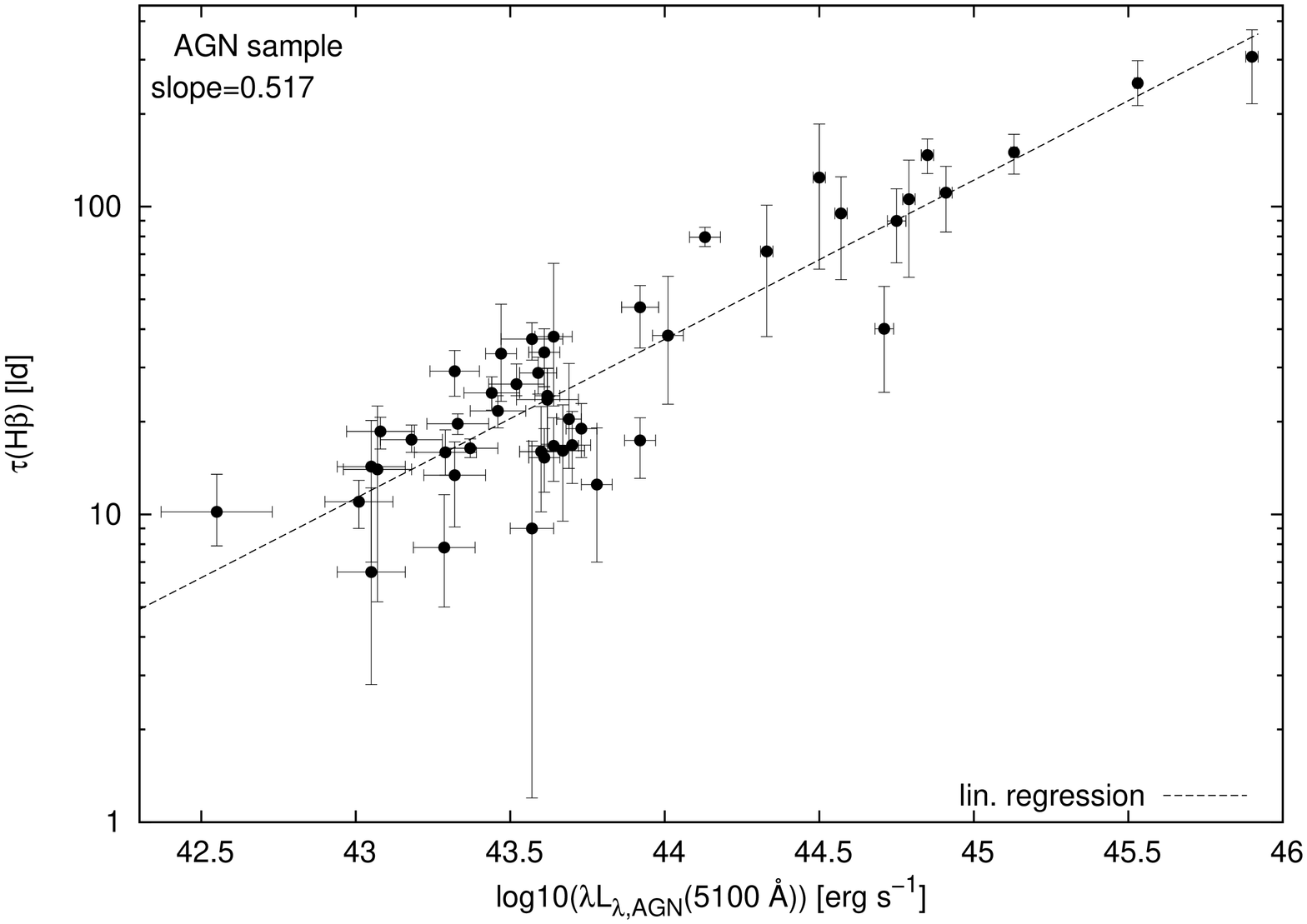} 
\includegraphics[width=6.0cm,angle=-90]{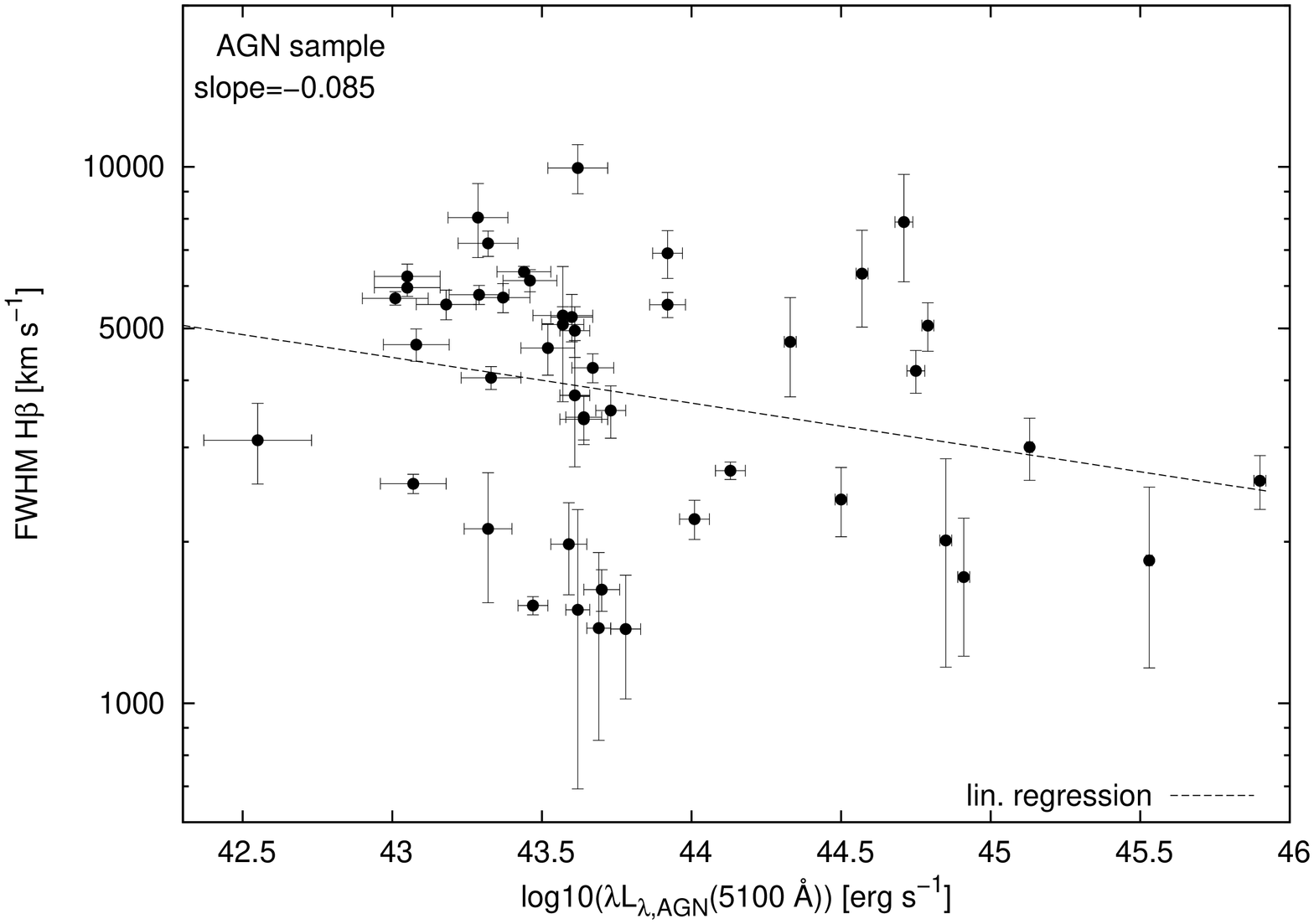} 
}
  \caption{Upper half: H$\beta$ BLR size–luminosity relation for the NGC~5548
data over 13 years 
(see Paper III) as well as the corresponding FWHM(H$\beta$)–luminosity
relation on the right side.
Lower half: same as upper half for all AGN in the Peterson et 
al.\cite{peterson04} sample (Paper I).}
   \label{corr_}
\end{figure*}
 \begin{table*}
 \caption{Correlation coefficients r (Pearson, Spearman, and  Kendall)
and probabilities P for random correlations for the H$\beta$ BLR size -
continuum luminosity relation as well as for the H$\beta$ FWHM -
continuum luminosity relation.}
    \centering
    \newcolumntype{d}{D{.}{.}{3}} 
\tabcolsep+5.3mm
    \begin{tabular}[H]{lrrrrdd}
 \htopline
  & $r_{p}$ & $r_{s}$  & $r_{k}$   & $P_{p}$&  \mcr{$P_{s}$} & \mcr{$P_{k}$} \\
 \hmidline
NGC 5548:  \Hb{} BLR size vs $\lambda{}L_{\lambda}$  & 0.743 & 0.773 & 0.600 & 0.009 & 0.015 & 0.010 \\
All AGN:  \Hb{} BLR size vs $\lambda{}L_{\lambda}$  & 0.901 & 0.774 & 0.613 & 0 & 1.556\times10^{-7}& 1.262\times{}10^{-9} \\
NGC 5548: FWHM(\Hb) vs $\lambda{}L_{\lambda}$ & -0.029 & -0.044 & -0.039 & 0.924 & 0.871 & 0.854 \\
All AGN:  FWHM(\Hb) vs $\lambda{}L_{\lambda}$ & -0.257 & -0.342 & -0.241 & 0.081 & 0.020 & 0.017 \\
\hbotline
 \end{tabular}
 \label{tab:stat}
 \end{table*}
%
Here we connected
the individual emission lines
 of the line-emitting regions of the narrow line galaxies NGC~7469
and NGC3783 and the line-emitting regions of the two variability
campaigns of NGC~5548. The line-emitting regions
 are drawn in
different colors: \ion{C}{iv}\,$\lambda 1549$ (blue), 
\ion{Si}{iv}+\ion{O}{iv]}\,$\lambda 1400$ (black), 
\ion{He}{ii}\,$\lambda 1640$ (cyan),
\ion{He}{ii}\,$\lambda 4686$ (green),
H$\beta$ (red).
%

There is a clear trend that the BLR geometries in AGN
are different for the individual galaxies as a function of their line widths.
They are not simple scaled-up versions 
depending only on the central black hole mass.
Besides the luminosity
there is at least one other significant parameter:
the rotational velocity of the central broad-line
region. This parameter controls the height of the line-emitting 
regions above the midplane.
Broader H$\beta$ and/or \ion{C}{iv}\,$\lambda 1550$ lines
scaled to the same Schwarzschild radius originate
closer to the midplane than do narrower species.

\subsection{Comparison of broad-line region
 geometries in different Seyfert galaxies scaled to
their optical luminosities.}

It has been shown before that the distances of the line-emitting regions
depend on the luminosities of the central ionizing sources
(e.g., Koratkar \& Gaskell\citealt{koratkar91a},
 Dietrich \& Kollatschny\citealt{dietrich95}, Kaspi et al.\citealt{kaspi00} 
Peterson et al.\citealt{peterson02}, \citealt{peterson04},
 Bentz et al.\citealt{bentz13}).
This BLR size–luminosity relation is shown separately for the H$\beta$ line
in NGC~5548 over 13 years (Fig.~16 upper left, Table~2 in this paper and
Table~1 in Paper III) and for the H$\beta$
line in all galaxies of the Peterson et al.\cite{peterson04} sample 
(Fig.~16 lower left,  Table~2 in this paper and supplementary information
in Paper I). 
 The continuum luminosities
at 5100~\AA{} given by Peterson et al.\cite{peterson04}
 have been corrected for the contribution of
their host galaxies (Bentz et al.\citealt{bentz13}).

A strong correlation is known to exist between the H$\beta$ BLR size and the
continuum luminosity in NGC~5548 for different variability campaigns as well
as for the AGN sample of the Peterson et al.\cite{peterson04}.
We calculated the corresponding
correlation coefficients (Pearson, Spearman, and  Kendall)
and probabilities for random correlations  (see Table~3).
More details regarding these coefficients can be found in
Kollatschny et al. \cite{kollatschny06}.
The slope $\alpha$ for the H$\beta$ data in NGC~5548 is
$\alpha$~=~0.49 in the H$\beta$ BLR size–luminosity diagram
(Fig.~16, upper left).
For these calculations
we did not take into account the two H$\beta$ measurements with very large
error bars ($\tau(H\beta)\le$10 days).
The slope has a value of
$\alpha$~=~0.517 when we consider all AGN in our sample (Fig.~16, lower left).


%
%
%
\begin{figure}
\includegraphics[width=8.0cm,angle=0]{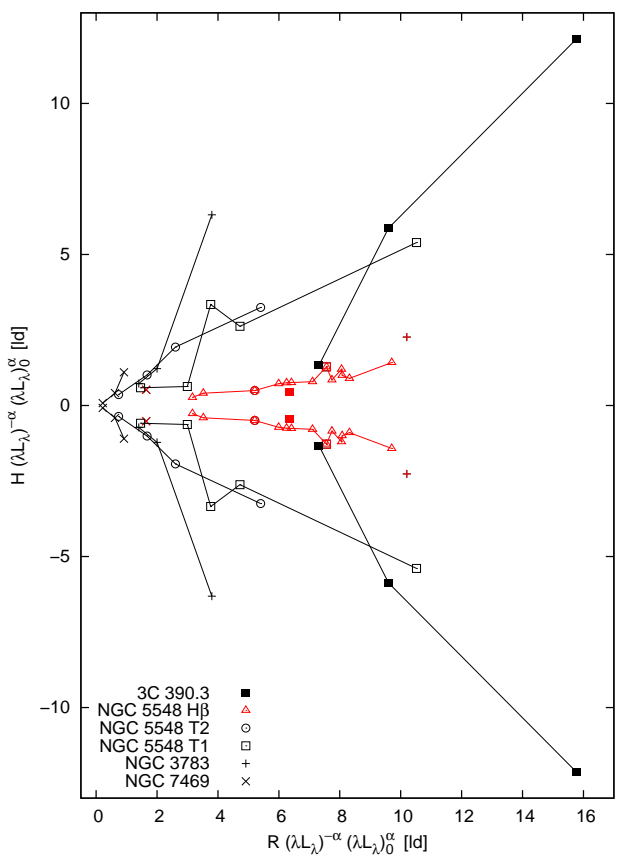} 
  \caption{Broad-line region structures in NGC~7469, NGC~3783, 
NGC~5548 (two epochs), and 3C~390.3
 scaled both to their individual continuum luminosities
$\lambda{}L_{\lambda}^{\alpha}$ at  5100 \AA{} and with respect
to the continuum luminosity $(\lambda{}L_{\lambda})_{0}^{\alpha}$ of
 NGC 3783 (based on corrected turbulent velocities $v_{\mathrm{turb}}$).
The emitting regions of the highly ionized lines of the individual galaxies are
connected by a solid line.
The H$\beta$ emitting regions are drawn in red. Those
of  NGC~5548 are connected by a solid red line.
}
   \label{disc_all_normlum.ps}
\end{figure}
%
%
\begin{figure}
\includegraphics[width=8.0cm,angle=0]{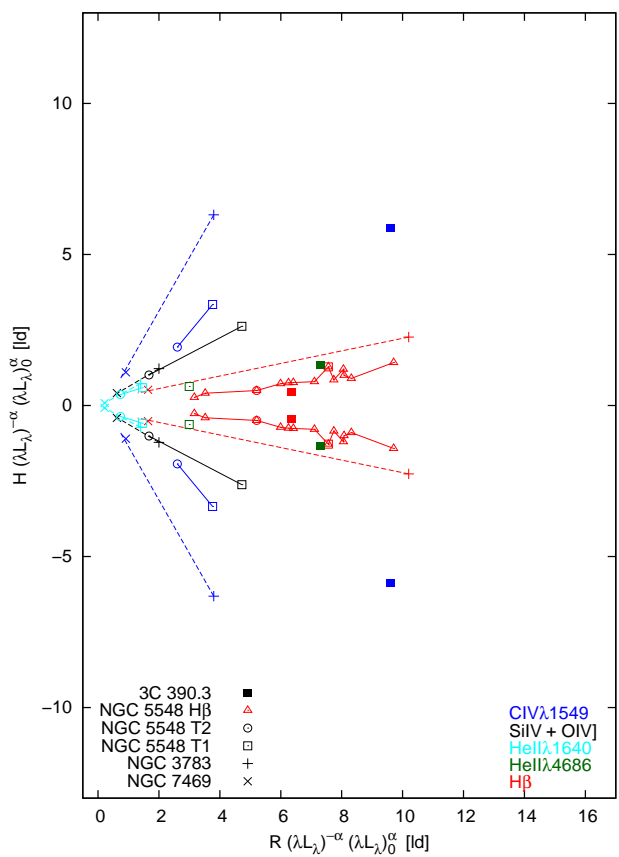} 
 \caption{Broad-line region structures in NGC~7469, NGC~3783, 
NGC~5548 (two epochs), and 3C~390.3
scaled to their individual continuum luminosities
$\lambda{}L_{\lambda}^{\alpha}$ at  5100 \AA{} (based on corrected
turbulent velocities $v_{\mathrm{turb}}$).
Same as Fig.~17, but the emitting regions of the different lines are
drawn in different colors: \ion{C}{iv}\,$\lambda 1549$ (blue), 
\ion{Si}{iv}+\ion{O}{iv]}\,$\lambda 1400$ (black), 
\ion{He}{ii}\,$\lambda 1640$ (cyan),
\ion{He}{ii}\,$\lambda 4686$ (green),
H$\beta$ (red).
The individual line-emitting regions of the two narrow
line Seyfert galaxies NGC~7469 and NGC~3783 are connected by a dashed
line to aid the eye. 
The line-emitting regions
 of NGC~5548 are connected by a solid line.}
\label{disc_all_same-ion3_normlum.ps}
\end{figure}
For comparison we show the FWHM of H$\beta$ as a function of their related
optical continuum luminosities at 5100 \AA{} on the right-hand side in  Fig.~16.
The FWHM are taken again from Peterson et al.\cite{peterson04} and
supplementary information in Paper I.
In that case there is only a weak
anticorrelation between the FWHM of H$\beta$
and the ionizing continuum luminosity
 with a large scatter. In both cases the slope  $\alpha$ is nearly flat:
the slope for the NGC~5548 data alone is  $\alpha$~=~-0.014, and 
we get a slope of  $\alpha$~=~-0.085 for the whole AGN sample.

Now we investigate the question whether BLR structures in AGN
are unambiguously characterized by their central luminosities,
which control the distances/radii of the line-emitting regions. 
We jointly present in  Figs.~17 and 18 the BLR structures of our four AGN
scaled both to their individual
continuum luminosities $\lambda{}L_{\lambda}^{\alpha}$ and with respect 
to the continuum luminosity $(\lambda{}L_{\lambda})_{0}^{\alpha}$ of NGC 3783.
For this scaling relation, we took the value of $\alpha$~=~0.533 (from Bentz et
 al.\citealt{bentz13}).
Again, these BLR structures are based on the observed high-ionization
as well as H$\beta$ emission lines.
The highly ionized lines (e.g., \ion{C}{iv}\,$\lambda 1549$)
are always emitted in a far more extended region above the midplane
in comparison to the H$\beta$ lines, as seen before in Figs 12 to 15.
The \ion{C}{iv}\,$\lambda 1549$
line showing the greatest scale height is also the line for which
Gaskell et al.\cite{gaskell07} found the
 highest covering factor,
 while they also found a relatively low covering factor
 for the Balmer lines.
 Reverberation mapping transfer functions of the low-ionization lines
(Krolik et al.\citealt{krolik91}; Horne et al,\citealt{horne91})
 are consistent with a flattened disk geometry as well.
In addition, one can see again that the BLR geometries in AGN
are different for the individual galaxies as a function
of their line widths (Fig.~18).
This means that besides the luminosity
there is at least one other significant parameter.
This parameter is the line width and/or
the rotational velocity of the central 
region and controls the height of the line-emitting 
regions above the midplane.
Broader H$\beta$ and/or \ion{C}{iv}\,$\lambda 1550$ lines
 originate
closer to the midplane than do narrower species.
We noticed this
 trend before when we scaled the broad-line region
structures with respect to their Schwarzschild masses 
and radii (Fig.~15).


\section{Conclusions}

We investigate the broad-line region structures of four AGN
that exhibit emission line profiles with
different widths: the H$\beta$ line widths FWHM range from 2000 to
10\,000  km\,s$^{-1}$.
The H$\beta$ lines are emitted in a more flattened configuration
above the midplane in comparison to the highly ionized lines. 
The H$\beta$ lines originate at heights
of 0.7 to 1.6 light-days only above the midplane
 at distances of 1.4 to 24 light-days.
The highly ionized lines originate at smaller radii than the H$\beta$ lines
and/or at greater distances from the midplane at $H/R$ values of 0.2 to
1.7. In total the emission lines do not originate in a thin
atmosphere of an accretion disk but rather in very extended regions
above the midplane, as  has been ascertained
from other lines of evidence (Gaskell \citealt{gaskell09}).
The observed geometries of the line-mitting regions resemble
the geometries of accretion disk wind models.
Furthermore, the angle of the central opening cone (generated by the
emitting regions of the highly ionized lines)
is small for the galaxies showing slow rotational velocities and increases
with the rotation velocity.

\begin{acknowledgements}
      Part of this work was supported by the German
      \emph{Deut\-sche For\-schungs\-ge\-mein\-schaft, DFG\/} project
      number Ko~857/32-1.
\end{acknowledgements}


\end{document}